\pdfoutput=1 

\documentclass[cits]{JINST}
\usepackage{amsmath}
\usepackage{ulem}

\title{Performance Study of Charcoal-based Radon Reduction Systems for Ultraclean Rare Event Detectors}

\author{M. Arthurs$^a$, D.Q. Huang$^a$, C.~S. Amarasinghe$^a$, E.~H. Miller$^b$ and W. Lorenzon$^a$
\thanks{Corresponding author.} \\
\llap{$^a$}Randall Laboratory of Physics, University of Michigan,
Ann Arbor, Michigan 48109-1040, USA \\
\llap{$^b$}SLAC National Accelerator Laboratory,
Menlo Park, California 94025-7015, USA \\

E-mail: \email{lorenzon@umich.edu}\\
}

\abstract{The continuous emanation of radon due to trace amounts of uranium and thorium in detector materials introduces radon to the active detection volume of low-background rare event search detectors. $^{222}$Rn produces a particularly problematic background in the physics region of interest by the ``naked'' beta decay of its $^{214}$Pb daughter nucleus. While charcoal-based adsorption traps are expected to be effective for radon reduction in auxiliary circulation loops that service the warm components of current {ton-scale} detectors at slow flow rates $(0.5-2\;SLPM)$, radon reduction in the entire circulation loop at high flow rates $\mathcal{O}({100s\;SLPM})$ is necessary to reach high sensitivity in future generation experiments. In this article we explore radon dynamics with a charcoal-based radon reduction system  in the main circulation loop of time projection chamber detectors. We find that even for perfect radon traps, circulation speeds of $2,000\;SLPM$ are needed to reduce radon concentration in a 10\,ton detector by 90\%. This is faster by a factor of four  than the highest circulation speeds currently achieved in dark matter detectors. We further find that the effectiveness of vacuum swing adsorption systems, which have been employed very successfully at reducing atmospheric radon levels in clean-rooms, is limited by the intrinsic radon activity of the charcoal adsorbent in ultra-low radon environments. Adsorbents with significantly lower intrinsic radon activity than in currently available activated charcoals would be necessary to build effective vacuum swing adsorption systems operated at room temperature for rare event search experiments. If such VSA systems are cooled to about $190\,K$, this requirement relaxes drastically.}


\begin{document}
\section{Introduction}
\label{section-introduction}

Radon is a radioactive noble gas that is re-supplied continuously from the decay chains of uranium and thorium present in practically every material of rare event detectors, and constitutes the dominant background source in many dark matter searches.  Because radon is an inert gas, it dissolves in noble liquid detectors and cannot be removed with high temperature getters.  Among the radon isotopes abundant in nature, $^{222}$Rn ($\tau = 5.516$ days), a progeny of $^{238}$U, is of particular concern.  The beta decay of its daughter $^{214}$Pb has a gammaless component (6\% b.r.) directly to the ground state of $^{214}$Bi. This ``naked'' beta decay can end up in the low-energy region of interest for dark matter searches, survive the nuclear recoil discrimination cut, and be indistinguishable from low-energy nuclear recoils of rare particle interactions in the active volume of the detector.  Discriminating against such background events is very challenging in the analysis.

Hardware mitigation is necessary to reduce the continuously re-supplied radon background for ton scale and larger noble-liquid rare event searches, including dark matter direct-detection experiments.  LZ\footnote{For illustration purposes we will occasionally refer to the LZ experiment, which is an experiment with a detector mass of about 10\,tons of LXe. Note, however, that the general arguments are not limited to one specific dark matter effort.} is one {ton-scale} dark matter experiment~\cite{c0} that addresses this need by introducing an in-line radon reduction system (iRRS) in an auxiliary circulation loop~\cite{c1}. XENONnT is a different {ton-scale} dark matter experiment that employs inline distillation columns to address this need~\cite{XnT}. The LZ iRRS, which is based on a \textit{single} adsorption trap, takes in a small stream $(0.5-1\;SLPM)$ of radon-rich gaseous xenon from the warm regions of the xenon gas circulation system, and returns the radon-reduced xenon back to the main circulation loop\footnote{The main circulation {loop} in a TPC detector refers to the gas circulation and purification system that is needed for TPC detectors to remove a) electronegative impurities, such as oxygen and water that limit the free electron lifetime and degrade the operation of the TPC, and b) radioactive noble gases. Purification from electronegative impurities is achieved with getters containing zirconium that are operated at high temperatures, typically near $600\,C$. This requires noble liquids to be gasified before they can be introduced to the getters. Since radioactive noble gases cannot be removed with those getters, radioactive noble elements have to be removed with other means.}. While it is expected to reduce an estimated $\mathcal{O}({20\;mBq})$ radon burden from the warm regions to below 1\,mBq, it does not have the capacity to purify the entire 10\, tons of liquid xenon.

For radon reduction in the entire system, rather than in a few select areas, an iRRS in the main xenon circulation loop becomes necessary. This requires a larger trap (i.e. more adsorbent) to accommodate the much higher flow rates needed for purifying multi-ton dark matter detectors. As explored in Sec.~\ref{section-CT}, scaling up charcoal based \textit{single-trap} radon reduction systems for multi-ton time projection chambers (TPCs) is impossible given the intrinsic radon emanation of currently-available charcoals, and impractical even if radon emanation were negligible.

Pressure swing adsorption (PSA) systems have been shown to be very effective at reducing atmospheric radon levels in clean-rooms~\cite{c2}. PSA systems are commonly employed as \textit{two-trap} systems where the main flow of the carrier gas is alternated between the two charcoal columns allowing one column to be filled while the other is purged. Pioneering the development of PSA technology for radon reduced clean rooms, vacuum swing adsorption (VSA) systems (where the purge is at pressures of $\mathcal{O}{(10\;mbar})$) have demonstrated radon reduction efficacy of $99.7\%$ in air at flow rates as high as $2,000\;SLPM$~\cite{c2,c3,c4}. Section~\ref{section-vsa} explores the effectiveness of a  swing adsorption system {suitable for noble liquid detectors that are operated at room temperature or cooled to almost noble liquid temperature.} The figures and simulations in this work are available from a public Gitlab repository~\cite{c8}.

\section{Radon Dynamics in a TPC Dark Matter Detector}
\label{section-rndyn}

\begin{figure}[ht]
\includegraphics[width=0.55\textwidth]{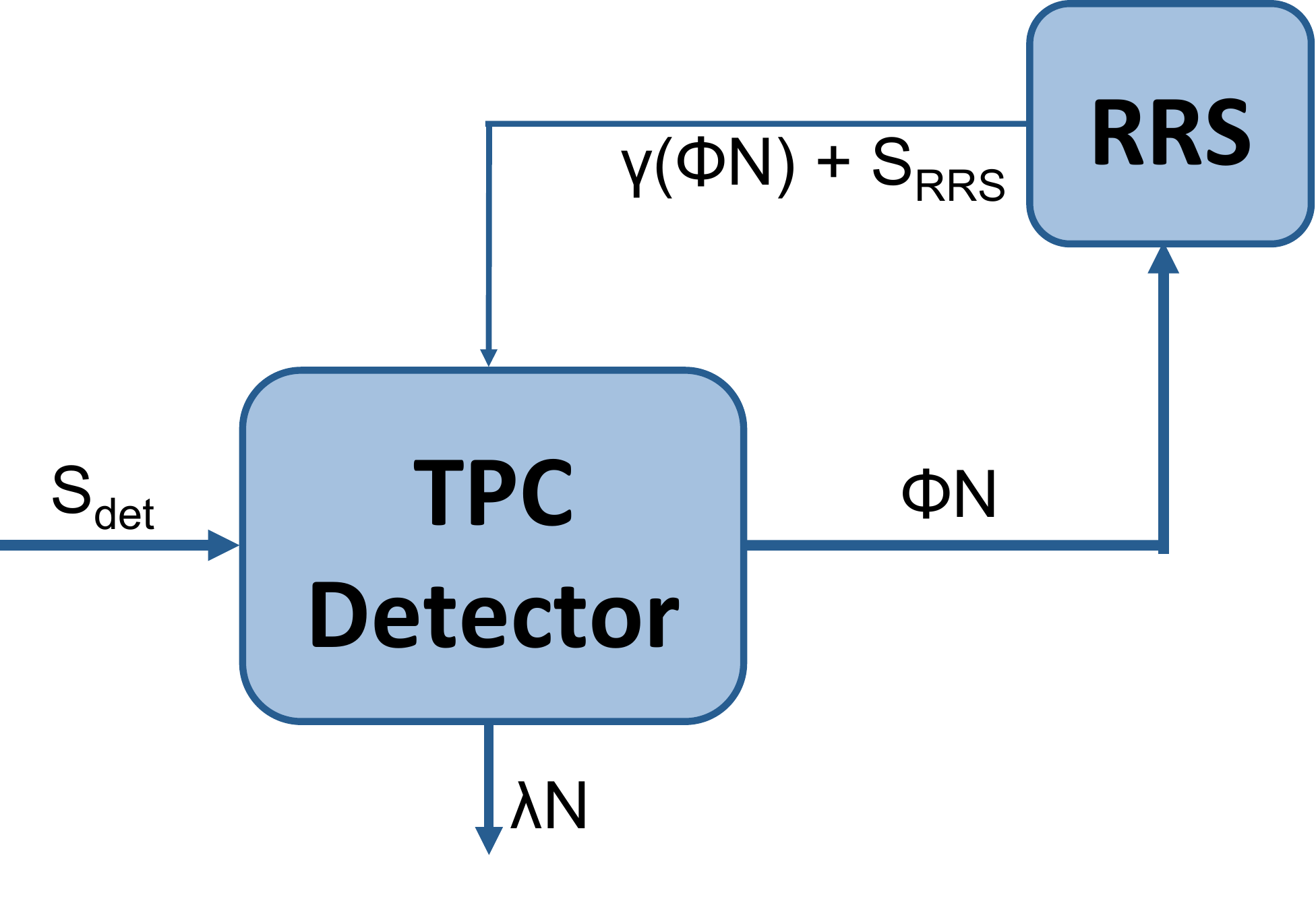}
\centering
\caption{Schematic diagram of radon dynamics in the active volume of a dark matter detector (TPC) with a radon reduction system (RRS) in the main circulation path. Note that $N$ is the number of radon atoms in the detector; $\Phi N$ is the rate of  radon atoms flowing out of the TPC; $\gamma$ is the fraction of {RRS inlet} radon atoms that escape the RRS; {$S_{RRS}$ is the radon activity due to the RRS.; and} $S_{det}$ is the radon activity in the detector.}
\label{f1}
\end{figure}

A schematic diagram of radon dynamics in a TPC detector with a RRS in the main circulation path is represented in Fig.~\ref{f1}. For a total radon emanation rate $S_{det}$ in the detector, the rate of change of the number of radon atoms in the detector, $N$, is given by
\begin{equation}\label{e1}
   \frac{dN}{dt} = -\lambda N - \Phi N + \gamma(\Phi N) + S_{RRS} + S_{det},
\end{equation}
where $-\lambda N = dN_{decay}/dt$ is the radon decay rate in the TPC with decay constant $\lambda  = 1/\tau $;  {$-\Phi N = dN_{TPCout}/dt$} is the rate of radon atoms flowing out of the TPC  set by the volume exchange time $T$ of the entire detector mass, with $T=1/\Phi$; $\gamma(\Phi N)$ is the inflow of radon atoms that escape the RRS, with $\gamma=N_{red}/N_{in}$ being the fraction of {external radon atoms introduced to the inlet of the trap ($N_{in}$) that escape} the RRS ($N_{red}$); and $S_{RRS}$ is the emanation rate from the RRS.  For simplicity, {sources within the circulation path other than the RRS have been excluded.}

Rearranging Eq.~(\ref{e1}) leads to

\begin{equation}\label{appendix_e2}
\begin{split}
    \frac{dN}{dt} = -\lambda N - \eta_{RRS}\Phi N + S_{RRS} + S_{det}= -\Lambda N + S,
\end{split}
\end{equation}
{where $\eta_{RRS}=1-\gamma$ is the remanent fraction of the RRS, referring to the fraction of inlet radon atoms that remain trapped in the RRS.} It can be solved to find the total number of radon atoms
\begin{equation}\label{e3}
    N(t) = -Ce^{-\Lambda t}+\frac{S}{\Lambda},
\end{equation}
where $\Lambda = \lambda  +\eta_{RRS}\Phi\,$ and $S=S_{det}+S_{RRS}$. In Eq.~(\ref{e3}), $C$ is an integration constant defined by the initial conditions.
{Since we are interested in the number of radon atoms in the TPC when equilibrium is reached, we can take the limit $t\rightarrow\infty$ to obtain the steady state number of radon atoms with a RRS in the main circulation path,}
\begin{equation}\label{e4}
    (N_{RRS})_{ss} = \frac{S}{\Lambda}  = \frac{S_{det}+S_{RRS}}{\lambda  +\Phi\eta_{RRS}}.
\end{equation}
{In the absence of a RRS, there will neither be radon reduction nor radon contribution from the RRS. The steady state radon count in a detector without a RRS in the main circulation path will be,
\begin{equation}\label{e5}
   ( N_{noRRS})_{ss} = \frac{S_{det}}{\lambda }.
\end{equation}
The fractional radon reduction with a RRS is expressed by the ratio of Eqs. (\ref{e4}) and (\ref{e5}), such that
\begin{equation}\label{e6}
   \left(\frac{N_{RRS}}{N_{noRRS}}\right)_{ss}  =\frac{T(1+S_{RRS}/S_{det})}{\eta_{RRS}\tau +T}.
\end{equation}
Hence, radon reduction efficacy with a RRS in the main circulation loop of the detector, defined as $\epsilon_{det}=1-(N_{RRS}/N_{noRRS})_{ss}$, becomes
\begin{equation}\label{e7}
    \epsilon_{det} = \frac{\eta_{RRS}\tau-T(S_{RRS}/S_{det}) }{ \tau\eta_{RRS}+T}.
\end{equation}}
{The detector efficacy describes the effective radon reduction in a TPC detector with a radon reduction system in the main circulation path.}
For a radon reduction system with no intrinsic activity, i.e. $S_{RRS}=0$, the radon reduction efficacy becomes
\begin{equation}\label{e7b}
    \epsilon_{det}= \frac{\tau }{\tau +T/\eta_{RRS}}.
\end{equation}
For a perfect radon reduction system
$(\eta_{RRS}=1)$, the highest achievable radon reduction efficacy
becomes
\begin{equation}\label{e8_}
    (\epsilon_{det})_{max}
    = \frac{\tau }{\tau +T}.
\end{equation}
This means that the maximum achievable radon reduction {efficacy} is ultimately limited by the {volume exchange time} of the detector.
\begin{figure}[ht]
  \centering
  \begin{minipage}[b]{0.45\textwidth}
    \includegraphics[width=\textwidth]{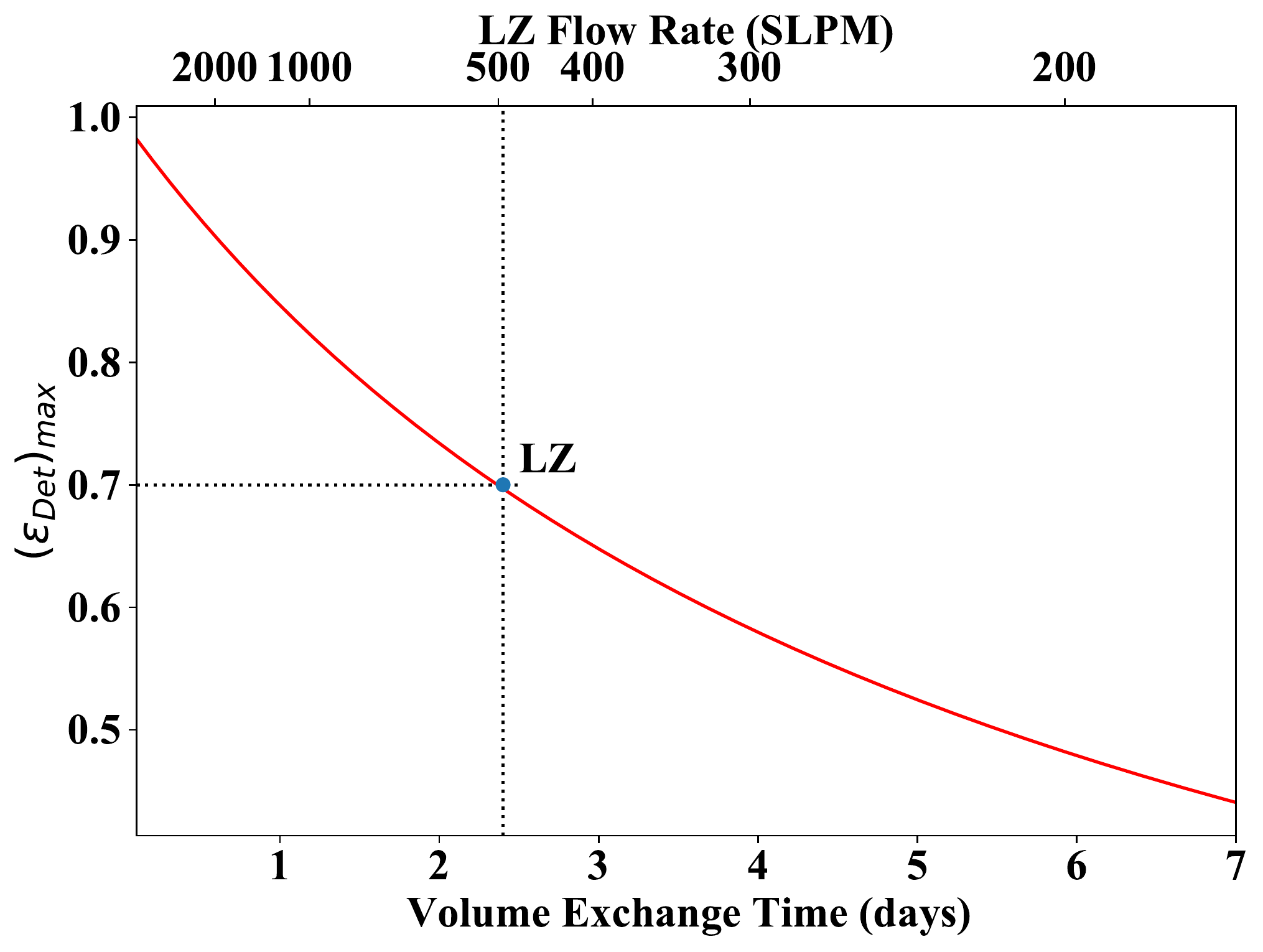}
  \end{minipage}\hspace*{5mm}
  \begin{minipage}[b]{0.45\textwidth}
    \includegraphics[width=\textwidth]{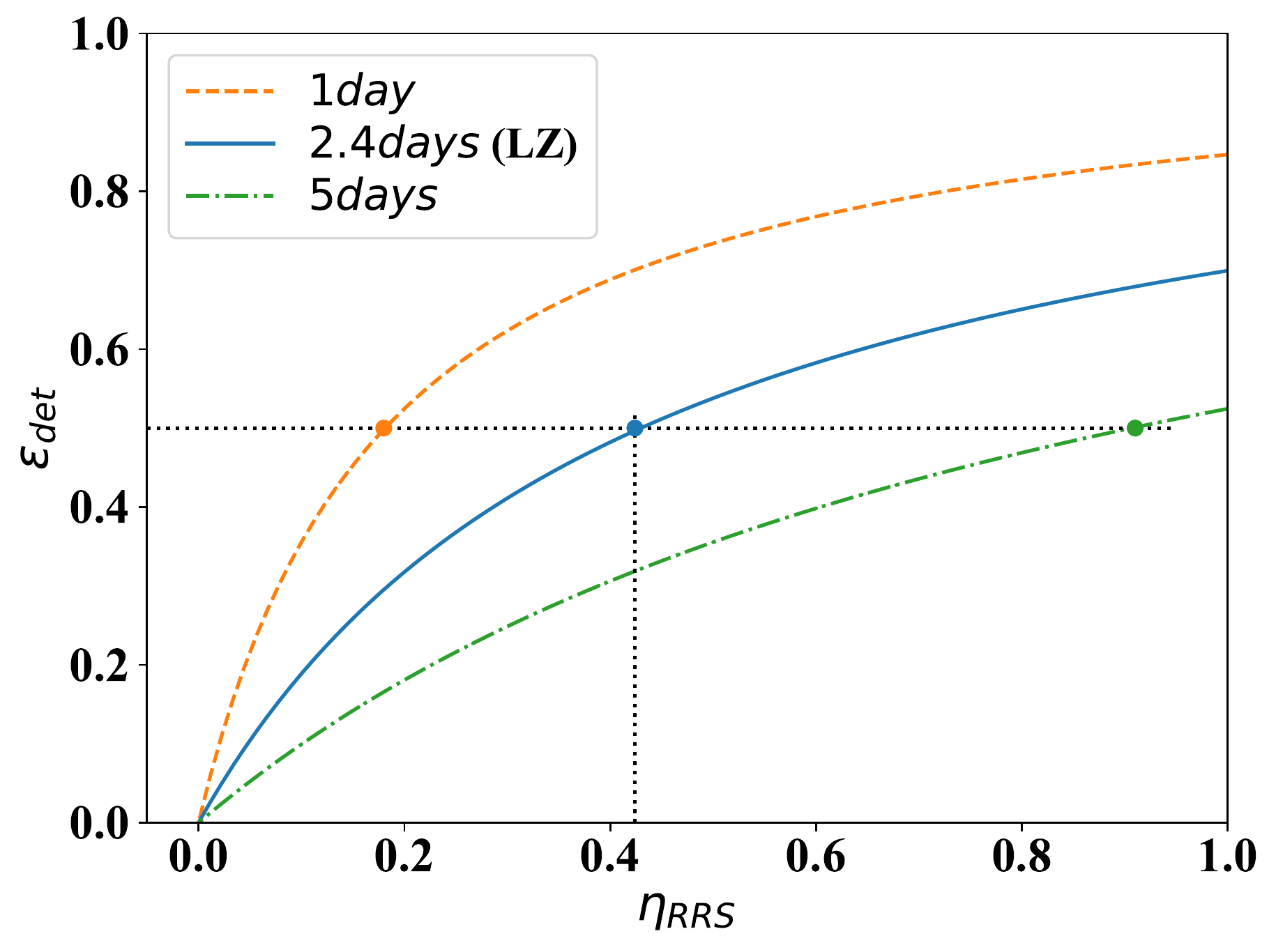}
  \end{minipage}
  \caption{Left panel: Maximum radon reduction {efficacy} achievable with a perfect radon trap in the main circulation loop  as a function of the {volume exchange time} of a TPC detector. The blue filled circle indicates the maximum efficacy for  a volume exchange time of 2.4 days, which is explored in more detail in the right panel. Also shown is the maximum efficacy as a function of the circulation flow rate specifically for a 10-ton detector (top horizontal scale). The dotted black lines indicate the maximum efficacy achievable for the conditions at LZ.  Right panel: Radon reduction efficacy achievable in a detector as a function of RRS {remanent fraction, assuming the intrinsic radon activity of the RRS is negligible,} for three different volume exchange times. The horizontal dotted black line indicates the conditions needed to achieve a factor of two radon reduction in {such a} detector.}
  \label{f2}
\end{figure}

The left panel of Fig.~\ref{f2} shows the maximum achievable radon reduction {efficacy} in a detector as a function of {volume exchange time} with a perfect radon reduction system in the main circulation path. Also shown in the figure is the maximum achievable radon reduction efficacy as a function of carrier gas circulation flow rate specifically for a 10-ton detector, such as in LZ. For that detector, with its  $F=500\;SLPM$ carrier gas {circulation} flow rate resulting in a volume exchange time\footnote{The volume exchange time is given by $T = M/\left(F\rho\right)$, where $F$ is the carrier gas {circulation} flow rate, $M$ is the total carrier gas mass, and $\rho$ is the carrier gas density {at STP}.} of about $T = 2.4$ days, and given the radon lifetime of $\tau  =  5.516$ days, at most a 70\% radon reduction efficacy (i.e. a radon reduction factor of 3.3) can be achieved. In order to reach radon reduction close to $90\%$ in LZ, flow rates of over $2,000\;SLPM$ are necessary. For such high flow rates it is very challenging if not impossible with current technology to use high-temperature getters for gas purification {(from electronegative impurities)}. Purification in the liquid phase using getters, operated at cryogenic temperatures, would have to be employed similar to what has been done in very large argon TPC experiments such as ICARUS~\cite{c7}.

The right panel of Fig.~\ref{f2} shows that for detectors with imperfect RRSs, the faster the volume exchange time, the lower the requirements on the RRS to achieve a certain radon reduction efficacy. If we declare that a successful RRS has to provide at least a factor of two radon reduction (i.e. $\epsilon_{det}=0.5$), volume exchange times of {at most} {$5.5\,$days} have to be achieved with a perfect RRS (shown in left panel). If shorter volume exchange times can be achieved, demands on the RRS can be significantly reduced (shown in the right panel). {For the conditions at LZ, a RRS with a 42\%  {remanent fraction} is sufficient to reach a 50\% efficacy in the detector. }

\section{Performance of a Single-Trap RRS}
\label{section-CT}

Radon reduction in the single-trap RRS approach is accomplished by maintaining radon breakthrough times that are long enough that the vast majority of the radon atoms entering the trap decay, while the carrier gas quickly traverses the trap.
The breakthrough time of a radon atom
in a charcoal trap, $t_b$, defined by the chromatographic plate adsorption model, is given by Ref.~\cite{c2} as
\begin{equation}
    t_b = \frac{mk_a}{f},
\label{Ae1}
\end{equation}
where $m$ is the charcoal mass, $k_a$ is the dynamic adsorption coefficient of radon on charcoal in a carrier gas, and $f$ is the volumetric flow rate of the carrier gas. This is an example of gas chromatography where one takes advantage of the different propagation speeds for radon and the carrier gas in the charcoal trap. The propagation speeds can vary by several orders of magnitude, particularly at cryogenic temperatures, where a ratio of $v_{Xe}/v_{Rn}=1,000$ has been reported~\cite{c6}. If the trap is large enough, so that radon needs a few lifetimes to reemerge on the other side of the trap, the overall radon concentration in the carrier gas is reduced accordingly.

{For a single trap with a breakthrough time $t_b$, the reduction of inlet radon atoms is given by an exponential decay law as}
\begin{equation}
    N_{red} = N_{in} e^{-\frac{t_b}{\tau }} =   N_{in} e^{-\frac{mk_a}{\tau f}} = N_{in} e^{-\frac{m}{\mu}},
\label{Ae2}
\end{equation}
where $N_{in}$ is the number of radon atoms that enter the trap and $N_{red}$ is the number of radon atoms that emerge from the trap, and $\mu=f\tau /k_a$ represents the characteristic mass of the trap, which is the mass of charcoal, at a given flow rate, needed to reduce radon activity by a factor of $e$. Remanent fraction of a single trap {(st)} is defined as
\begin{equation}
    \eta_{st} = 1- \frac{N_{red}}{N_{in}} =
    1-  e^{-\frac{mk_a}{\tau f}} = 1-  e^{-\frac{m}{\mu}},
\label{Ae2b}
\end{equation}
Additionally, Equation~(\ref{Ae2}) can also be expressed in terms of input activity $A_{in}$ and {reduced} activity $A_{red}$, since  $A = N/\tau$, so that
\begin{equation}
    A_{red} = A_{in} e^{-\frac{t_b}{\tau }} =   A_{in} e^{-\frac{mk_a}{\tau f}} = A_{in} e^{-\frac{m}{\mu}}.
\label{Ae2a}
\end{equation}

\begin{figure}[ht]
\centering
\includegraphics[width=0.5\textwidth]{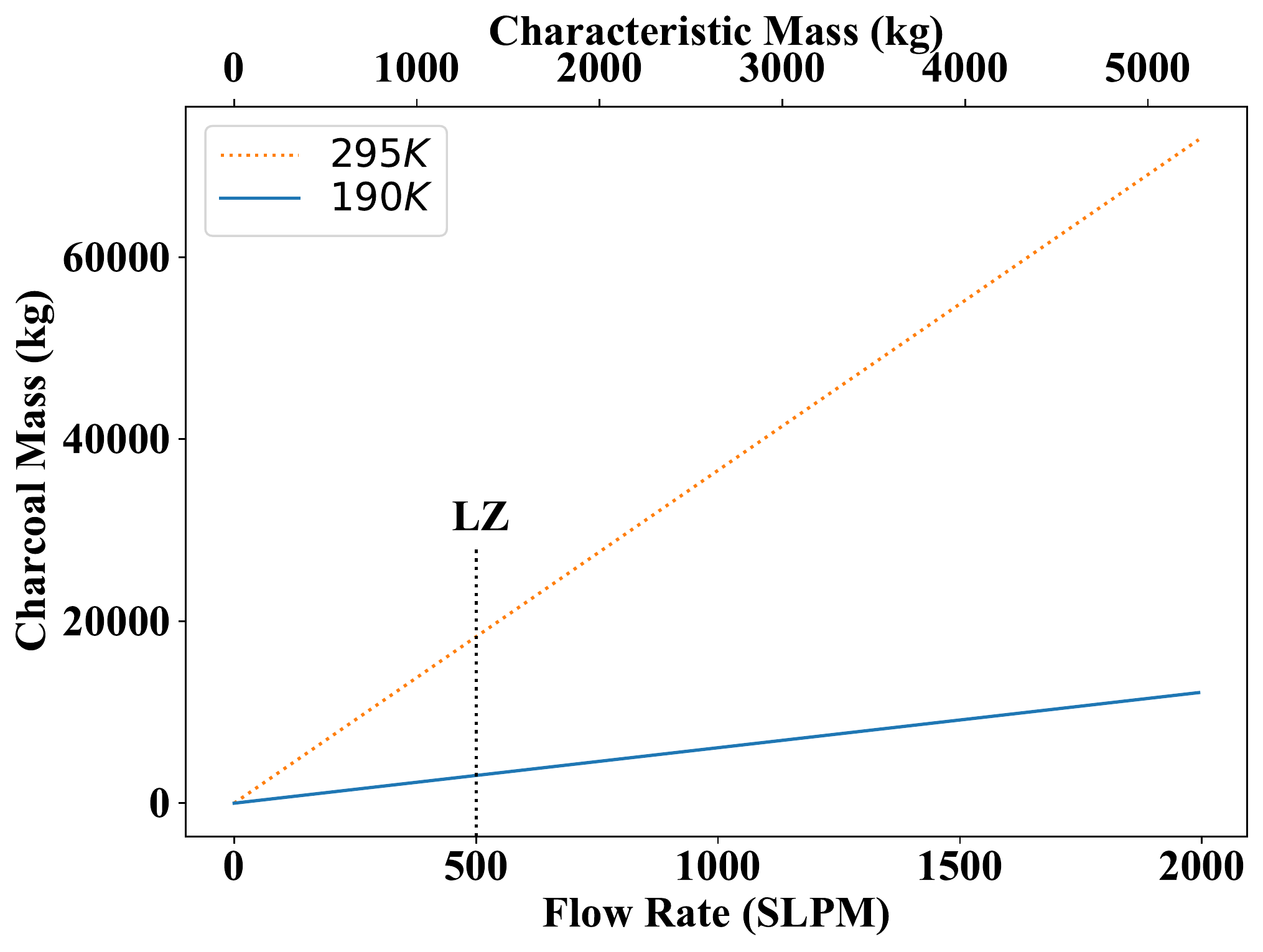}
\caption{The amount of charcoal needed  for 90\% {remanent fraction} as a function of carrier gas flow rate for a single-trap RRS with zero intrinsic radon activity. The {blue} solid line is at 190\,K $(k_a=3,000\,l/kg)$, and the  {orange} dotted line is at room temperature, 295\,K $(k_a=500\,l/kg)$. The dotted black line indicates the $500\;SLPM$ carrier gas circulation flow rate in LZ.}
\label{f3}
\end{figure}

For charcoals with the same adsorption properties as that used in the LZ iRRS~\cite{c1}  (such as Saratech charcoal with $k_a=500\,l/kg$ at 295\,K, and $k_a=3,000\,l/kg$ at 195\,K) the amount necessary to achieve 90\% {remanent fraction} as a function of the {circulation} flow rate of the carrier gas is shown in Fig.~\ref{f3}. {The figure demonstrates that for the $500\;SLPM$ carrier gas circulation flow rate} at LZ, it would take about $3,000\,kg$ of charcoal at $190\,K$ {or} $20,000\,kg$ at $295\,K$ to achieve a 90\% {remanent fraction}. A $3,000\,kg$ cold trap of charcoal with a density of about $0.6\,g/cm^3$ would occupy a volume of roughly $5\,m^3$, and adsorb almost $5,000\,kg$ of  xenon~\cite{c1}. Thus, scaling of single traps to sustain the high flow rates needed for multi-ton dark matter experiments is not a viable option, not even for ideal traps.\\[-8mm]

For a realistic trap, charcoal has intrinsic activity that contributes radon atoms to the output of the trap. Intrinsic activity of a charcoal is typically given by its specific activity $s_o$ in units of $mBq/kg$. For a charcoal trap of mass $m$ with specific activity $s_o$, the total radon activity of the trap will be $ms_o$ (i.e. the number of radon atoms emanating from the total charcoal mass per second). Note that not all of the emanated radon atoms escape the trap {\textemdash} some of them decay in the trap. For a charcoal trap of a mass $M$ {$(M \propto v_{Rn} t_b)$}, assuming uniform radon emanation over the entire trap, {the emanation of an infinitesimal charcoal slice of mass $dm$ is given by}
\begin{equation}
    s_odm = \frac{A_{em}}{M}dm = A_{em}\frac{dm}{M} = A_{em}\frac{v_{Rn}dt}{v_{Rn}t_b}
    =A_{em}\frac{dt}{t_b},
\label{e3aa}
\end{equation}
where $A_{em} = s_oM$ is the total activity of a {trap of mass $M$ with specific activity $s_o$. Emanated radon from that slice decays while it travels through the trap.} The radon contribution at the output of the trap from a such infinitesimal slice is given by
\begin{equation}
    dA_{st} = \frac{A_{em}}{t_b}e^{-\frac{t}{\tau }}dt.
\label{e3a}
\end{equation}
{{Emanated} radon atoms from a slice at the beginning of the column have to travel through the entire column and thus need the full radon breakthrough time to reach the end, while radon atoms from a slice close to the end leave the column immediately. Therefore, an integration of Eq.~(\ref{e3a}) from $t=0$ to $t=t_b$ yields the total radon contribution of the trap}
\begin{equation}
    A_{st} = \frac{s_om}{t_b}\int_{0}^{t_b}e^{-\frac{t}{\tau }}dt = s_om\frac{\tau}{t_b}\left(1-e^{-\frac{t_b}{\tau }}\right) = s_of\frac{\tau}{k_a}\left(1-e^{-\frac{mk_a}{\tau f}}\right).
\label{Ae6}
\end{equation}
{In order to find the effective output of a single trap, Eqs.~(\ref{Ae2a}) and (\ref{Ae6}) can be {added}, to give the total activity at the output of a single trap as}
\begin{equation}
\begin{split}
    A_{out} & = A_{in}e^{-\frac{mk_a}{f\tau }} + s_of\frac{\tau }{k_a}\left(1-e^{-\frac{mk_a}{f\tau }}\right), \\ & = A_{in}e^{-\frac{m}{\mu}} + s_o\mu\left(1-e^{-\frac{m}{\mu}}\right).
\label{e8}
\end{split}
\end{equation}
Note that for sufficiently large traps, where $m\gg \mu$, the lowest achievable radon activity at the output of the trap is given by $A_{out} \approx s_o \mu$,  and thus depends on the specific activity but not on the total mass of the charcoal.

\subsection{Single-trap RRS with constant radon inlet}\label{subsubsection-const-inlet-st}

{The performance of {a single-trap radon reduction system with constant radon inlet} can be explored in terms of efficacy, which encapsulates both the reduction of external radon introduced to the inlet of the trap, and radon emanation from the trap due to its intrinsic activity. Efficacy describes the net fraction of radon atoms removed by the trap, such that a fraction of 1 indicates a perfect trap, i.e. no radon atoms emerge from the trap; a fraction of 0 indicates an ineffectual trap, i.e. the same number of radon atoms enter and exit the trap; and a {negative fraction indicates a  {harmful} trap, i.e. more radon atoms leave than enter the trap.}}

The single-trap efficacy, defined as $\epsilon_{st}=1-A_{out}/A_{in}$, can be expressed in terms of Eq.~(\ref{e8}), as
\begin{equation}\label{e11}
    \epsilon_{st}=1-A_{out}/A_{in} = 1- e^{-\frac{m}{\mu}} - \frac{s_o \mu}{A_{in}} \left(1-e^{-\frac{m}{\mu}}\right) = \bigg[1-\frac{s_o \mu}{A_{in}}\bigg]\left(1-e^{-\frac{m}{\mu}}\right).
\end{equation}
It increases with increasing input radon activity. This makes the technique particularly well-suited for radon reduction from radon-rich environments.

The relevant parameter for the efficacy of a single-trap RRS is the ratio of the specific activity of the charcoal and the input radon activity, $(s_o /A_{in})$. Together with the characteristic mass of the charcoal, it determines the maximal efficacy of a trap in the limit of $m\rightarrow\infty$ as
\begin{equation}\label{e12}
   \left(\epsilon_{st}\right)_{max} = 1-\frac{s_o }{A_{in}} \mu =1-\left(\frac{s_o}{A_{in}}\right)\left(\frac{f\tau} {k_a}\right) = 1 - \frac{f}{f_{critical}},
\end{equation}
where $f_{critical} = (A_{in}/s_o) (k_a/\tau)$ is the critical flow rate of the carrier gas for a given single-trap condition ($A_{in}$, $s_o$, and $k_{a}$). Note that at the critical flow rate the efficacy of the trap becomes zero, and above it the trap becomes harmful. In order to design an effective single trap it is necessary to chose a charcoal with high adsorptive properties (large $k_a$)  and low intrinsic activity (small $s_o$), which translates to maximizing the critical flow rate of a trap.

\begin{figure}[ht]
\centering
\includegraphics[width=0.5\textwidth]{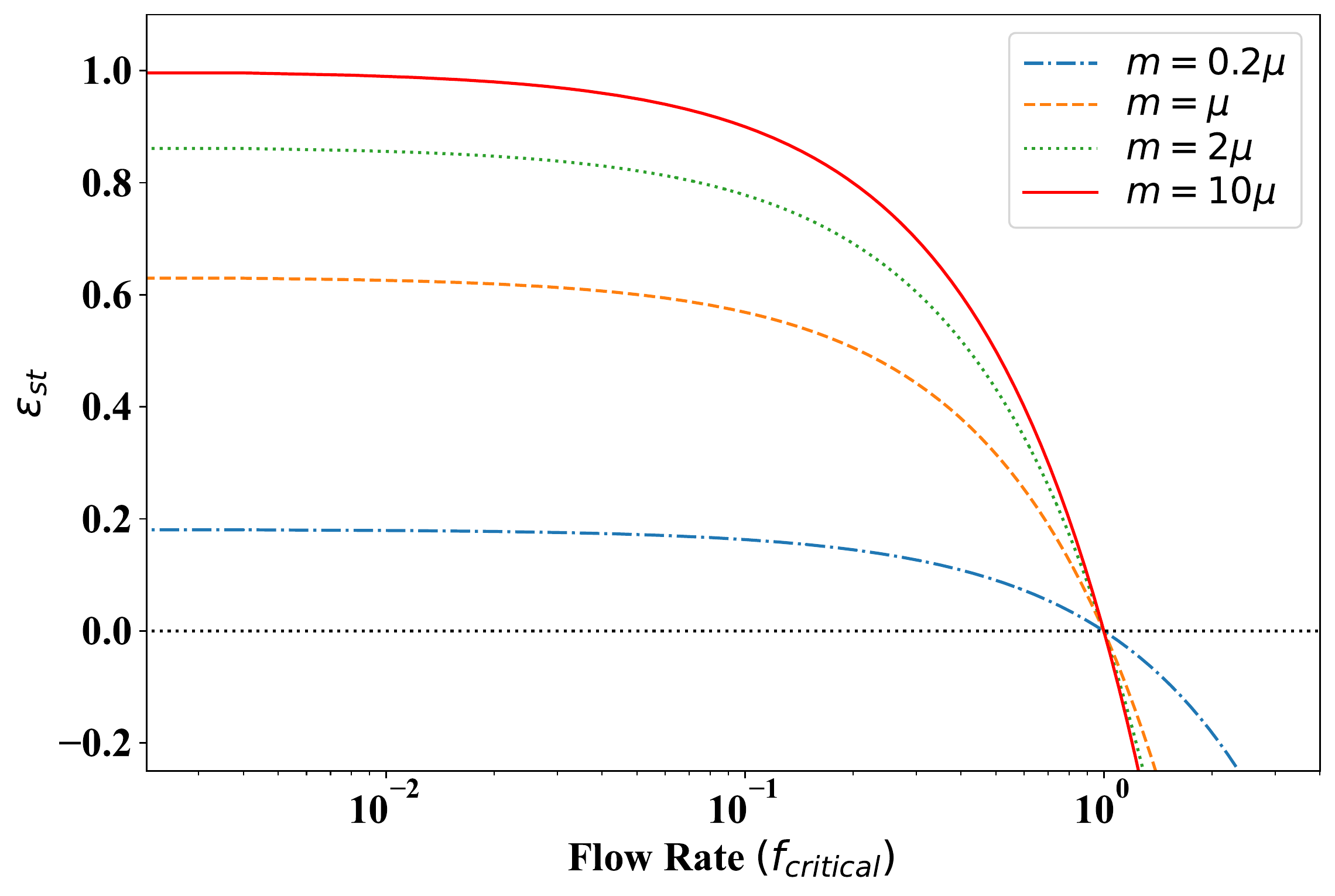}
\caption{Efficacy of a single-trap RRS as a function of the flow rate in units of $f_{critical}$. At the critical flow rate, the efficacy becomes zero, independent of trap mass. Above it,  the trap becomes harmful. The various curves represent different trap masses in units of the characteristic mass $\mu$.}
\label{cflow_vs_eff}
\end{figure}

Figure~\ref{cflow_vs_eff} explores the dependence of the single-trap efficacy on the flow rate for various charcoal masses. The flow rate is shown in units of critical flow rate, and the mass is given in units of characteristic charcoal mass of the trap. The figure shows that independent of charcoal mass, carrier gas flow rates of about an order of magnitude lower than the critical flow rate are necessary to reach maximal efficacy. At the critical flow rate, trap efficacy becomes zero independent of charcoal mass.

\begin{figure}[h]
\centering
\includegraphics[width=0.5\textwidth]{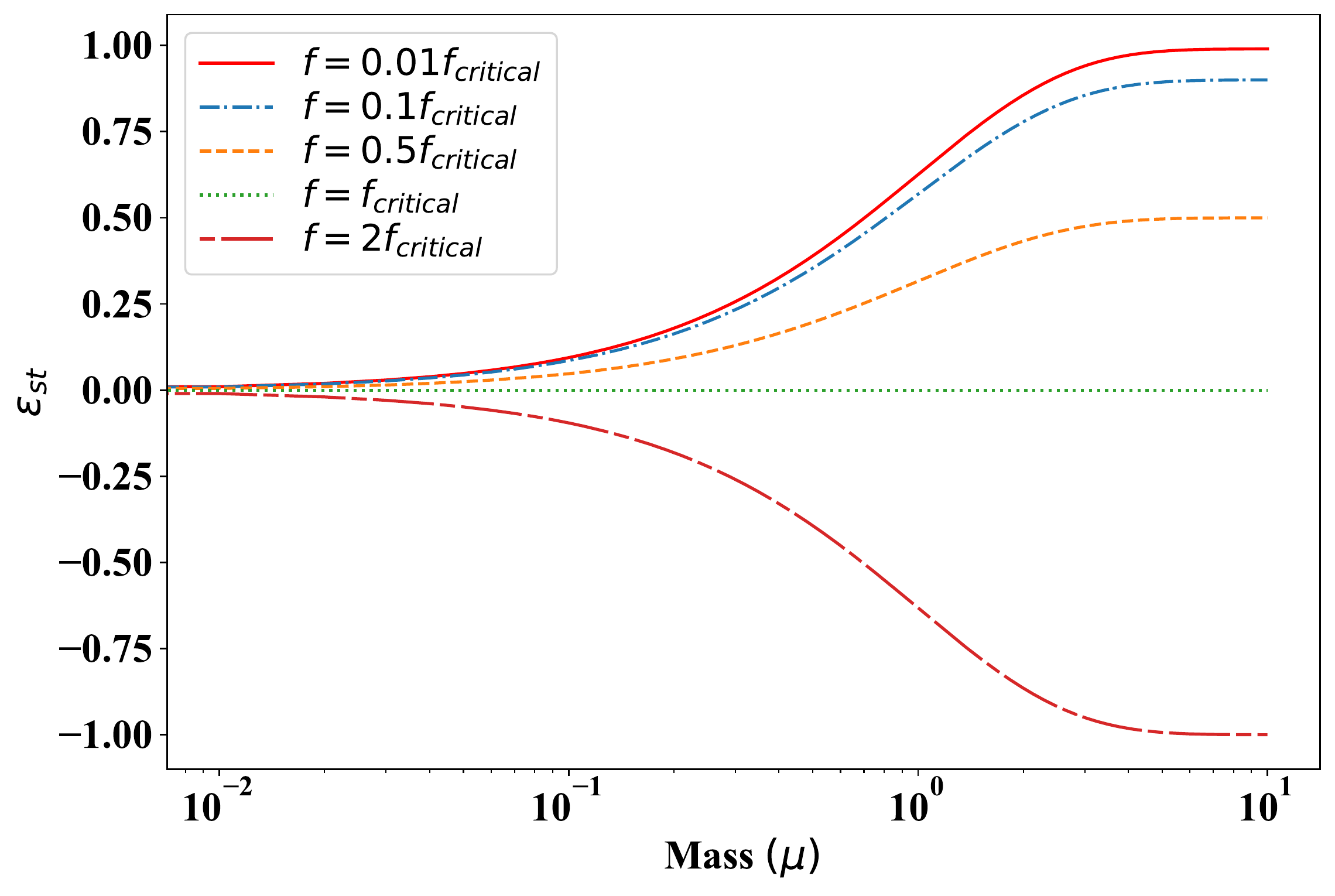}
\caption{Efficacy of a single-trap RRS as a function of characteristic mass $\mu$. For trap masses above about four characteristic masses $(4\mu)$, the increase in the efficacy is asymptotically small. The various colors represent different flow rates relative to the critical flow rate $f_{critical}.$
}
\label{cmass_vs_eff}
\end{figure}

Figure~\ref{cmass_vs_eff} shows the single-trap efficacy as a function of trap mass for various flow rates of the carrier gas. The mass is given in units of characteristic charcoal mass of the trap, and flow rate is shown in units of critical flow rate. For flow rates below the critical flow rate, the efficacy of the trap increases with increasing charcoal mass and rapidly approaches its maximal value, until it reaches a trap mass of $\mathcal{O}(4\mu)$, above  which the increase in the efficacy is asymptotically small. {For flow rates above the critical flow rate, the efficacy of the trap becomes increasingly more negative with increasing charcoal mass until it approaches its maximal negative value for a trap mass of $\mathcal{O}(4\mu)$.}

\begin{figure}[ht]
\centering
\includegraphics[width=0.5\textwidth]{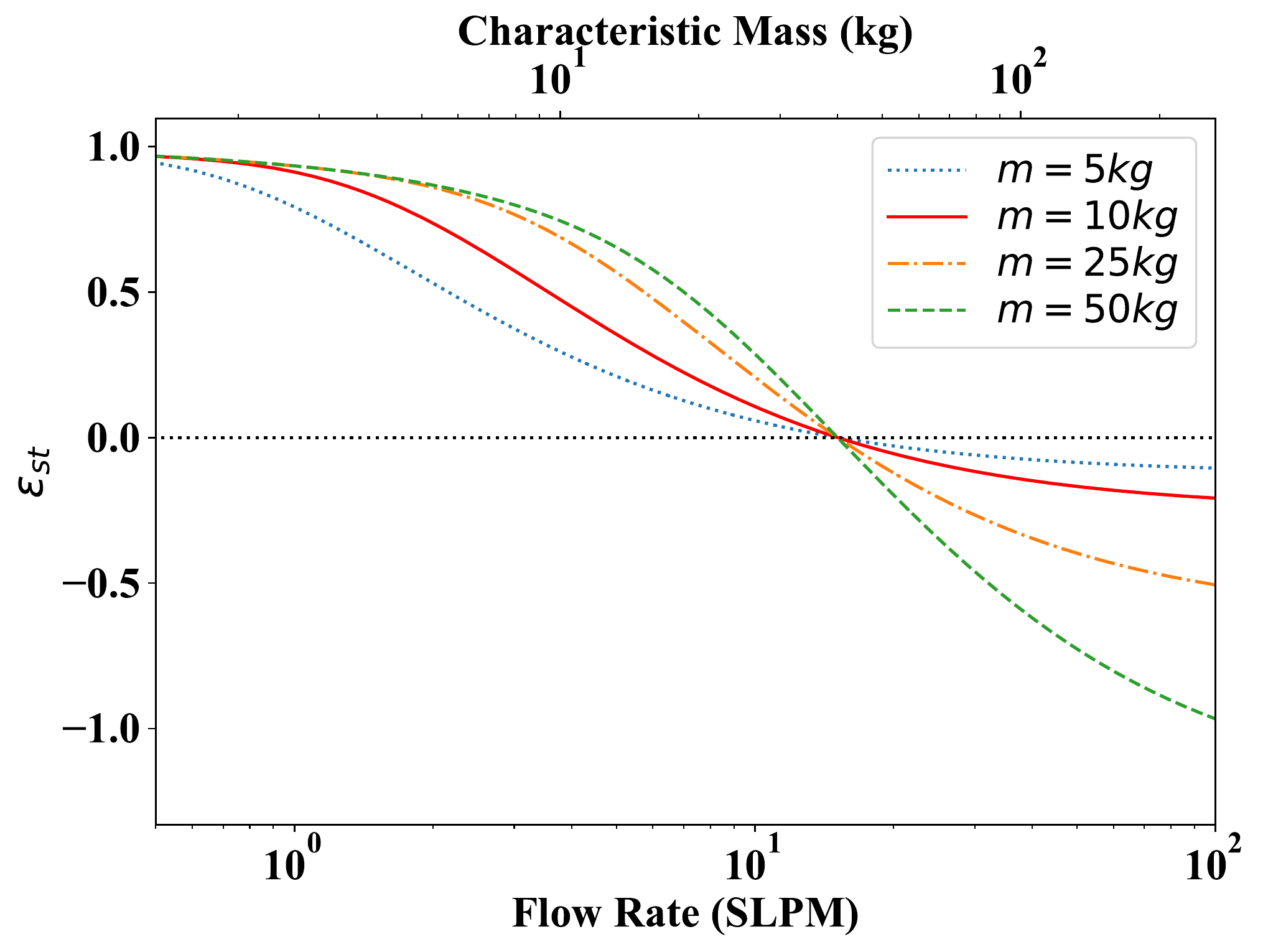}
\caption{Efficacy of LZ iRRS (a $10\,kg$ charcoal single trap with dynamic adsorption coefficient of $k_a=3,000\,l/kg$ (at $190\,K$), total inlet radon activity of $A_{in} = 20\,mBq$) as a function of the flow rate of the carrier gas. The various curves represent  efficacies for different  masses of charcoal.}
\label{LZ_singleTrap}
\end{figure}

\noindent
For illustration purposes, let us explore the trap efficacy for the LZ iRRS, shown in Fig.~\ref{LZ_singleTrap}, which employs a synthetic charcoal called Saratech~\cite{c1} in a single-trap approach. Saratech has a dynamic adsorption coefficient of $500\,l/kg$ at room temperature that increases to $3,000\,l/kg$ as the temperature falls to $190\,K$, which is slightly above the liquefaction temperature of xenon. Running the trap at cryogenic temperatures is advantageous, as it requires relatively small amounts of charcoal. The charcoal used in the LZ iRRS has an intrinsic activity of $\mathcal{O}({0.5\;mBq/kg})$~\cite{c1}. For an inlet radon activity of $20\,mBq$, the critical flow rate is  $f_{critical}  \approx 15\,SLPM$. For flow rates below the critical flow rate, where radon contribution from the charcoal is smaller than radon reduction due to adsorption, a greater mass of charcoal results in a higher efficacy, as shown in Fig.~\ref{LZ_singleTrap}. At the critical flow rate, shown as the inflection point in Fig.~\ref{LZ_singleTrap}, the radon contribution from the charcoal compensates the reduction due to adsorption.  Above the critical flow rate, the efficacy becomes negative indicating that the radon reduction system becomes harmful and introduces more radon atoms to the detector than it removes. Figure~\ref{LZ_singleTrap} also shows that at the relatively low flow rates of $0.5-1\;SLPM$ at LZ, radon reduction efficacies of more than 90\%  can be achieved with a $10\,kg$ (i.e. $7.6\,\mu$) charcoal trap.

\section{Swing Adsorption for Radon Reduction}
\label{section-vsa}

Vacuum swing adsorption (VSA) systems have been developed for radon reduction in clean rooms for flow rates of order $1,000\;SLPM$. This is in contrast to single-trap radon reduction systems whose performance is set by the steady-state radon output, which limits the flow rate. VSA systems are {systems} {commonly} consisting of two charcoal columns where the flow direction of the carrier gas is periodically switched between the columns.

\begin{figure}[ht]
\includegraphics[scale=0.99]{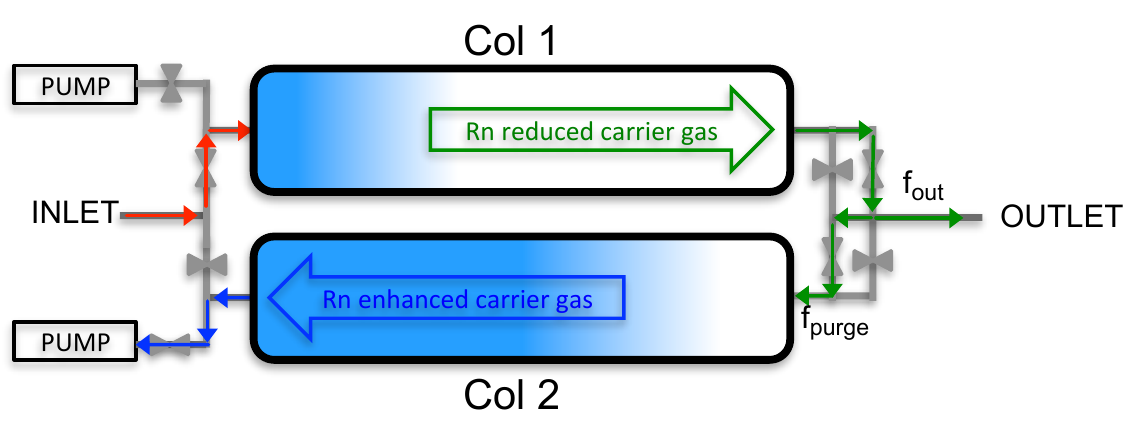}
\centering
\caption{A schematic view of a VSA system for radon reduction in air. Flow of the input air is alternated between columns 1 and 2 to prevent radon escaping from the outlet ends of the columns. While one column is fed with air, the other is purged with a small stream of radon-reduced air. The shades of blue indicate the radon concentrations in the two columns, the red arrows  indicate the flow of the input gas, the green arrows indicate the flow of the radon-reduced output gas, and blue arrows indicate the pump-out flow.}
\label{f8}
\end{figure}

A schematic view of a VSA system for radon reduction in air is presented in Fig.~\ref{f8}. Ambient air is fed into column 1 (here the feed column) for a time much shorter than the time required for radon atoms to transit the column, while column 2 (here the purge column) is purged {at low pressure} with a small stream of radon-reduced air from the outlet end of column 1\footnote{Typically about 10\% of the radon-reduced air emerging from the outlet of the feed column is used for purging the radon-enhanced column while a vacuum pump maintains the column pressure at around {10\,mbar}.} to flush the radon atoms out.
{As described in Section \ref{subsubsection-real},} { the low pressure of the purge is necessary to obtain a regeneration cycle that is faster than the feed cycle.  }
At the end of this cycle, column 2 has been regenerated and is ready to be fed with outside air, while column 1 has accumulated radon and is ready to be purged. With the beginning of the next cycle, the outside air is directed into column 2 (now the feed column), while column 1 (now the purge column) is purged. By the end of the second cycle, each column has gone through one feed and one purge cycle. The time required to complete these two cycles is typically called a swing-cycle period.

By switching a given flow between the two columns, each column may be much smaller than the column in a single-trap RRS. Unlike in single-trap reduction systems, where radon atoms are retained in the charcoal for many lifetimes, in a swing system they are flushed out of the system. Additionally, unlike single-trap reduction systems, which are typically cooled down to cryogenic temperatures, VSA systems are shown to reach high efficacy even at room temperature. Over the past decade, VSA technology has been improved to reach radon reduction efficacy in air of greater than $99.9\%$~\cite{c3,c4}.
For an inlet radon activity of about $80\,Bq/m^3$, reduction factors of greater than 1,000 were achieved, reducing the clean room radon activity down below the sensitivity of the RAD7 measurement device, with an upper limit of $0.067\,Bq/m^3$~\cite{c4}.

\subsection{Feasibility of Swing Adsorption RRS for Xenon}\label{subsection-vsa}

Considering the great success of VSA systems for radon-reduced clean rooms, we will now explore the viability of such a system for full scale radon reduction in a rare-event TPC detector, taking into account some distinct differences.

Since the radon content introduced to a VSA system due to the intrinsic activity of charcoal is typically much smaller than that in atmospheric air, it is mostly ignored in VSA systems used for radon reduction in clean rooms.
Conversely, in a liquid xenon dark matter detector with a radon content as low as 1\,atom/kg of xenon, the introduction of a charcoal trap could very well introduce more radon than it removes.   For simplicity, we will start with ignoring intrinsic activity (Sec.~\ref{subsubsection-quasi-ideal}), and then study the impact of non-zero intrinsic activity on VSA systems (Sec.~\ref{subsubsection-real}).

\begin{figure}[ht]
\includegraphics[width=0.75\textwidth]{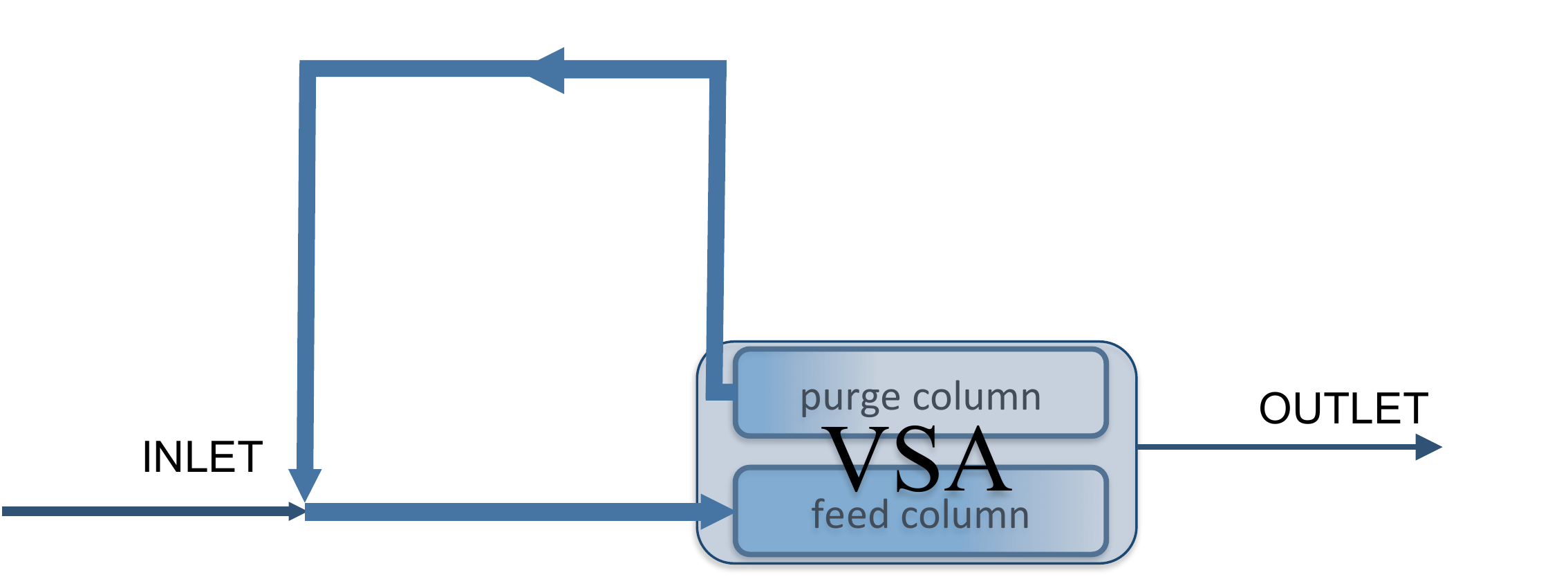}\centering
\centering
\caption{A schematic view of a VSA system with a feedback loop for radon reduction in xenon. Flow of the input xenon is alternated between the two columns,  but unlike in a radon reduction system for air, where the purged air is released back into the atmosphere, the purged xenon is returned through a feedback loop to the inlet of the VSA. {Note that the purge gas in the feedback loop is flowing in the  direction indicated by the arrows.}
}
\label{f9}
\end{figure}

Furthermore, in contrast to air purification systems, where the purged air is released back into the atmosphere, xenon is expensive, and needs to be captured and returned to the purification system as shown schematically in Fig.~\ref{f9}.  Therefore, rather than pumping and releasing the xenon gas into atmosphere, the radon-rich purge gas has to be returned to the inlet of the swing system.  In such a system, the radon atoms become effectively trapped and accumulate in the feedback loop. Accumulation of radon atoms in the feedback loop continues until it is balanced by the decay of the radon atoms and steady state is reached. Note that such a system conveniently provides a mechanism for radon atoms to decay outside of the TPC detector.

Because of the cyclic nature of the swing system, its columns never reach steady state. The full modeling of the system is therefore more involved and must track radon concentrations throughout each column and propagate them over time.  The exact behavior will depend strongly on the choice of charcoal~\cite{c1}, the geometry of the columns, the pumping speed of the system, and other system-dependent properties. While this system-dependent modeling is beyond the scope of this work, models prepared for other systems have shown that radon appearing at the VSA output is primarily due to the long diffusive tail of the radon front as it propagates through the charcoal~\cite{c2}. Thus, the remanent fraction of the VSA will depend on the elution curve of radon in the trap as well relative values of the cycle time and the trap breakthrough time. For simplicity, we fold this into a single constant remanent fraction for the feed column when modeling the performance of the VSA.

\subsubsection{Swing Adsorption RRS with Feedback Loop and zero Intrinsic Activity}\label{subsubsection-quasi-ideal}

In order to evaluate steady state conditions for a VSA system with a feedback loop, let us consider the dynamics of radon atoms in a single cycle.  For simplicity, and to compare with radon reduction systems in clean rooms, we will start with ignoring intrinsic radon activity. Let us consider the situation where some radon atoms entering at the inlet are allowed to escape the VSA system.

{In this specific model the relevant parameters are $\eta_{feed}$, $r_{purge}$, and $t_{feed}$. The feed column remanent fraction, $\eta_{feed}$, represents the fraction of input radon atoms being trapped in the feed column of the VSA; $t_{feed}$ is the time that a VSA column is in the feed stage, which must be less than the breakthrough time, $t_b$, of the column to minimize radon atoms escaping from that column;  $r_{purge}$ is the fraction of the radon-reduced carrier gas that is used for purging the radon-enhanced purge column; and $r_{out} = 1 - r_{purge}$ is the fraction of the radon-reduced carrier gas that flows back to the TPC detector.} {Unlike for a single-trap, where the remanent fraction for a given breakthrough time is defined according to Eq.~\ref{Ae2b},} {this model does not provide a direct relationship between the remanent fraction of the feed column and its breakthrough time. The challenge is that the feed column remanent fraction depends not only on the breakthrough time but also on the  particular shape of the elution curve. Including this in the model would be beyond the scope of this paper. However, it requires making assumptions about the magnitude of the remanent fraction.}

\noindent
{The evolution of radon atoms after the $n^{th}$ feed of the VSA is given by}
\begin{subequations}
\begin{equation}
    (N_{out})_{n} = (N_{in})_{n}(1-\eta_{feed}){e^{-t_{feed}/\tau}}r_{out},
\label{e20N}
\end{equation}
\begin{equation}
  (N_{loop})_{n} = (N_{in})_n(1-\eta_{feed}){e^{-t_{feed}/\tau}}r_{purge} +(N_{in})_{n}\eta_{feed}e^{-t_{feed}/\tau},
\label{s20L}
\end{equation}
\begin{equation}
  (N_{in})_{n+1} = N_{det} + (N_{loop})_{n},
\label{s20N}
\end{equation}
\end{subequations}
where $N_{out}$ is the number of radon atoms that flow back into the TPC detector, $N_{in}$  is the number radon atoms that enter the feed column, which includes both the constant supply from the detector, $N_{det}$, as well as the radon atoms from the feedback loop. The {first} term in Eq.~(\ref{s20L}) represents the {number of radon atoms} that escaped the feed column and are reintroduced into the purge column by the purging gas. The last term in Eq.~(\ref{s20L}) represents {the radon atoms in the purge column that were trapped during the previous cycle when that column was in the feed stage. Eq.~(\ref{s20N})} { gives the number of inlet radon atoms of the next cycle.}
{Given the $\mathcal{O}{(10\;mbar)}$ pressure in the purge column, the breakthrough time of the purge column is much {shorter} than the breakthrough time of the feed column~\cite{c5b}.} This means that the time radon atoms spend in the purge column is much shorter than the time radon atoms spend in the feed column. {For simplicity}, the decay of radon atoms in the purge column is therefore considered in the following feed cycle. 

The characteristics of Saratech charcoal are used to determine the appropriate breakthrough times for the VSA columns. The elution curve of Saratech charcoal for radon in xenon {(or argon)} carrier gas lacks significant tails in the front or the back, {and has a much smaller width than elution curves of other charcoals.} {The elution curve for a $70\,g$ trap with a $200\,$min breakthrough time, shown in Ref.~\cite{c1}}, { indicates that no significant amount of radon escapes during the first $100\,$min, or 50\%, of the mean breakthrough time. Furthermore, effectively all the radon escapes within $100\,$min after the mean breakthrough time.} By increasing the trap size to $\mathcal{O}{(20\;kg)}$, the smallest of the column sizes discussed here,  while maintaining the same aspect ratio and the same $200\,$min breakthrough time, the gas velocity $v$ increases by a factor of {6.6}\footnote{In order to maintain the same breakthrough time, the carrier gas velocity must scale with the column length which, with a constant length/diameter ratio, increases as the cube root of the mass.  The carrier gas velocity of a $20\,kg$ trap compared to a $70\,g$ trap then scales by $\left(\frac{20,000\,g}{70\,g}\right)^{1/3}=6.6$.} and the longitudinal diffusion, which scales as $\sqrt{v^{-1}}$~\cite{c10}, will decrease by a factor of {2.6}. Therefore it is expected that, for a $20\,kg$ trap, the radon will transit within about 20\% of the breakthrough time.  For a $100\,kg$ trap, the radon transit time will be about 15\% of the breakthrough time. For this reason, these traps have breakthrough times only 50\% longer than the feed cycle time, yet negligible breakthrough is expected.

\begin{figure}[ht]
  \centering
  \begin{minipage}[b]{0.47\textwidth}
    \includegraphics[width=\textwidth]{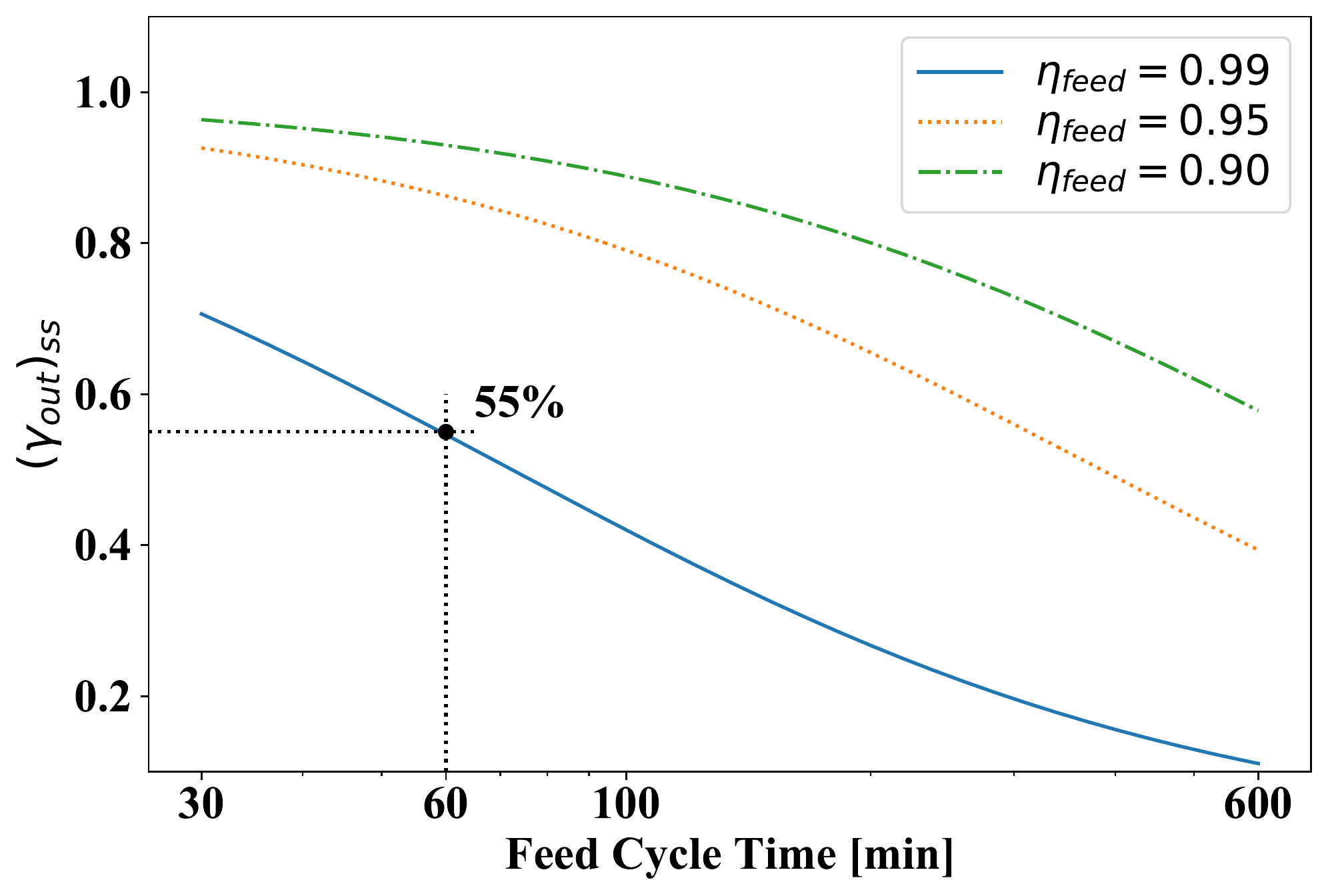}
  \end{minipage}\hspace*{5mm}
  \begin{minipage}[b]{0.435\textwidth}
    \includegraphics[width=\textwidth]{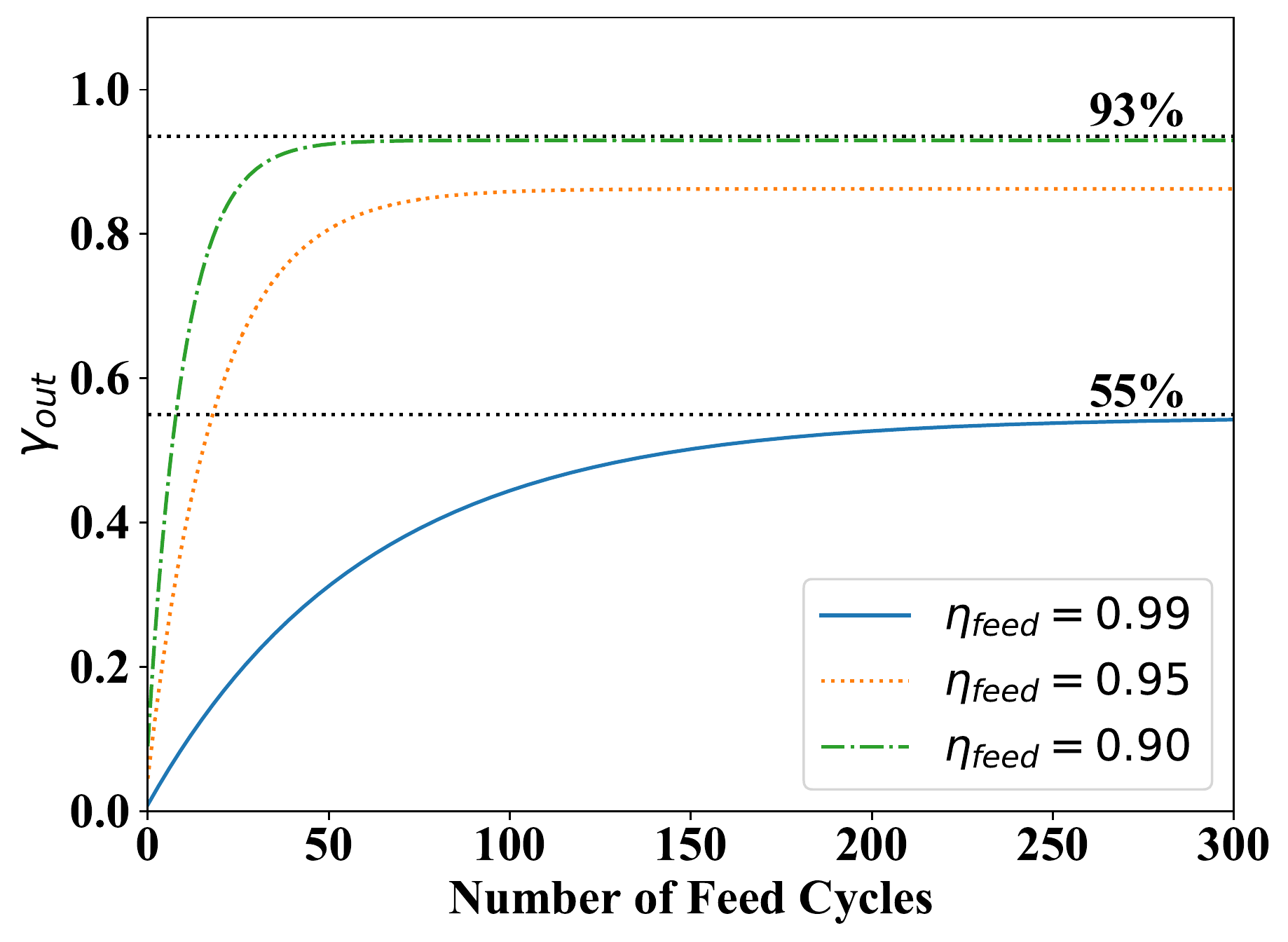}
  \end{minipage}
  \caption{Dynamics of radon atoms in a VSA system with 10\% purge flow fraction in the feedback loop for feed column remanent fractions of 99\%, 95\%, and 90\%.  The intrinsic activity of charcoal is ignored in this computation.   Left panel: steady state fraction of radon atoms escaping the VSA versus feed cycle times for xenon purification in $30-600\,$min range.
  Right panel: fraction of radon atoms escaping the trap versus the number of feed cycles for a feed cycle time of $60\,$min {and three feed column remanent fractions. As indicated  with the horizontal dotted lines, it takes about 300 (60) feed cycles to approach a steady state output radon fraction of 55\% (93\%) in the VSA with feed column remanent fractions of 99\% (90\%).}
  }
  \label{f55}
\end{figure}

As an example, the left panel of Fig.~\ref{f55} illustrates the steady-state output radon fraction, $(\gamma_{out})_{ss}$, for  VSA remanent fractions of 99\%\footnote{For a feed column with remanent fraction of 99\% , 1\% of the radon atoms entering the column are allowed to escape it, while the other 99\% remain in the column {during the feed cycle}.}, 95\%, 90\% with a 10\% purge flow fraction in the range of feed cycle times $30\,$min to $600\,$min. {Note that the performance of such a RRS, which has negligible intrinsic activity, does not depend on the number of radon atoms supplied by the detector, $N_{det}$. Therefore, its performance is expressed in terms of radon fraction, $\gamma_{out} = N_{out}/N_{det}$ and $\gamma_{in} = N_{in}/N_{det}$, rather than in terms of radon atoms.} The right panel of Fig.~\ref{f55} demonstrates that up to 300 feed cycles are necessary to reach steady state, $(\gamma_{out})_{ss} =0.55$, for a feed cycle time of $60\,$min.  {Details for why a $60\,$min feed cycle time is used for the VSA models are discussed in Appendix~\ref{appendix_B}.} Although not explicitly shown here, fewer feed cycles are needed to reach steady state as the feed cycle times get longer.  In order to increase the steady state radon {remanent fraction} in a VSA system, defined as $\eta_{RRS} = 1- \gamma_{ss}$, one can increase the feed cycle time. This requires very large charcoal columns, since the breakthrough time, which must be larger than the feed cycle time, grows linearly with charcoal mass. This does not only increase the cost associated with the increased trap size, but also the amount of xenon stored in the charcoal, which is about $0.4\,kg/kg$ at room temperature {and $1\,atm$}~\cite{c1}, and can become a significant fraction of the entire xenon mass.

Note that the assumption of a feed column with remanent fraction of 99\% may be optimistic.  Relaxing that number to 90\%  will increase the steady state  output radon fraction from 55\% to 93\%.  This relaxation may be necessary for charcoal beds where long, non-Gaussian tails at the front of their elution curves provide significant radon leakage even for feed cycle times much shorter than the mean breakthrough time. As described in Sec.~\ref{subsubsection-quasi-ideal}, this can be avoided by careful selection of the charcoal, and is less significant with larger charcoal columns. {It may therefore not be too optimistic to {increase the remanent fraction from 90\% to} 95\% for VSA systems that contain Saratech charcoal, {which has no long, non-Gaussian tails at the front of its elution curve.}}

\subsubsection{Adding a cold single trap to the Feedback Loop} \label{trap-in-loopl}

\begin{figure}[ht]
\includegraphics[width=0.7\textwidth]{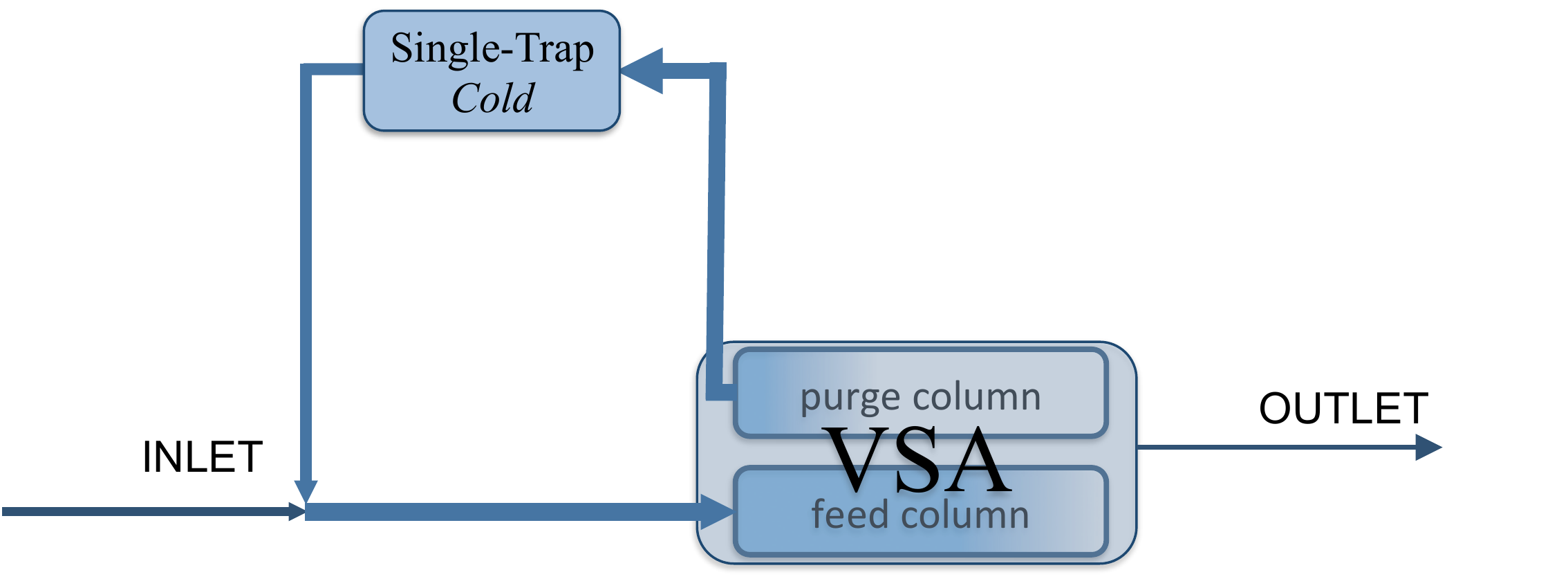}
\centering
\caption{A schematic of a VSA radon reduction system with a single, cold trap in the feedback loop for radon reduction in xenon. The cold trap greatly enhances the efficacy of the VSA system. {The arrows indicate the direction of the purge gas flowing through the feedback loop and the single trap.}}
\label{f11}
\end{figure}

An improvement is to integrate a single-trap RRS, which is preferably cooled, in the feedback loop of the VSA system, shown schematically in Fig.~\ref{f11}, such that the radon-enhanced gas from the purge column passes through the single-trap RRS before it is fed back into the inlet of the VSA. Such a trap provides a space for radon atoms to decay before being returned to the VSA. {In such a system, only a small fraction of the entire carrier gas circulation volume has to pass through the single-trap RRS, allowing the singe-trap to be relatively small.} In addition, we will show that this trap can have a relatively low {remanent fraction} while still significantly improving the performance of the system.

The addition of a single trap in the VSA feedback loop can be implemented in the radon dynamics model with a small modification in Eq.~(\ref{s20L}), so that
\begin{equation}
  (N_{loop})_{n} = (1-\eta_{st})\left[(N_{in})_n(1-\eta_{feed})r_{purge} +(N_{in})_{n}\eta_{feed}\right]{e^{-t_{feed}/\tau}}.
\label{e21N}
\end{equation}
where {$\eta_{st}$ is the remanent fraction} of the single trap. The inclusion of this trap also smooths out variations in radon concentration, which would otherwise be greater at the beginning of the cycle than at the end, justifying the approximation that $\eta_{feed}$ is constant over the course of a cycle.

\begin{figure}[ht]
\includegraphics[scale=0.5]{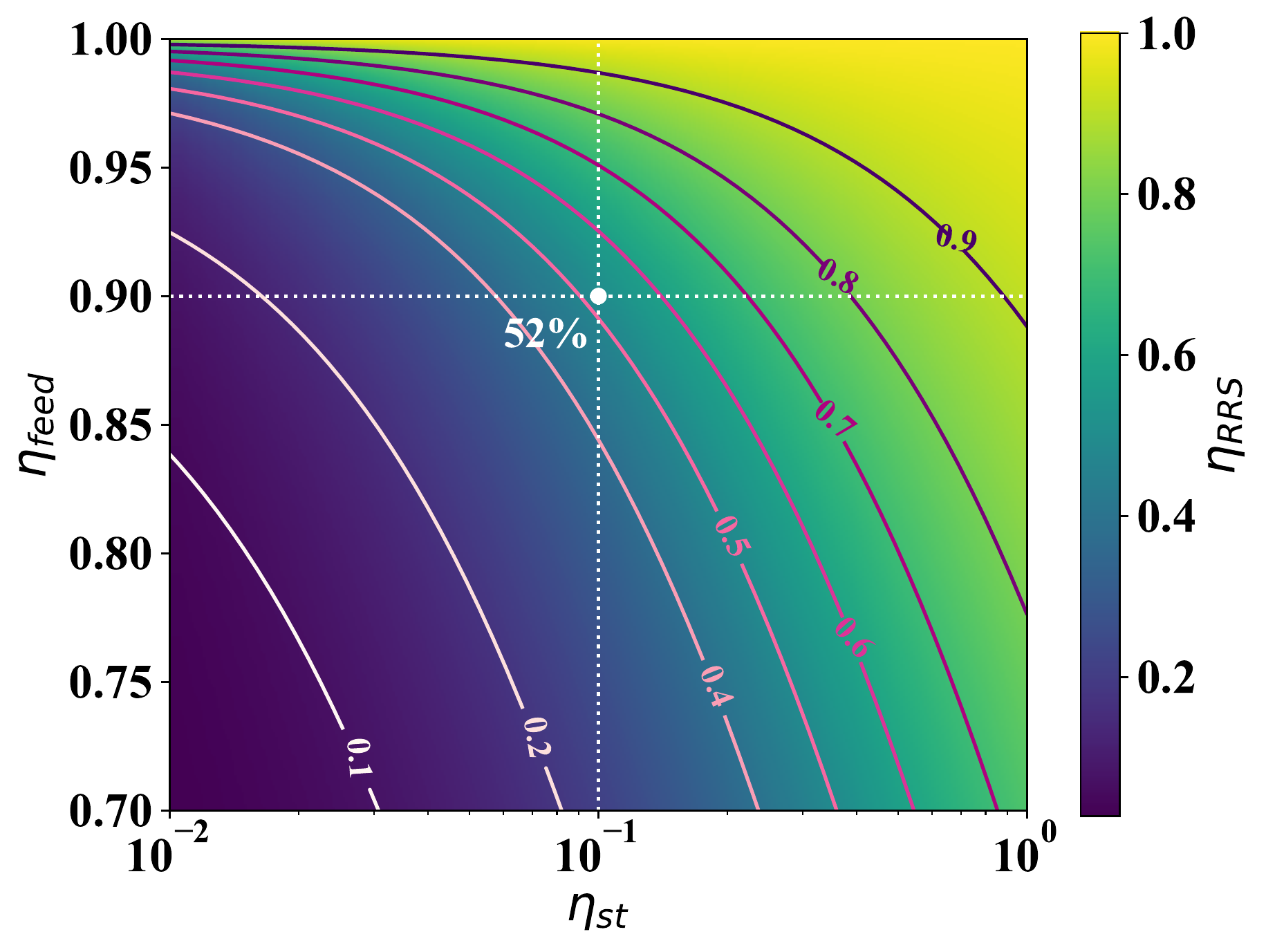}
\centering
\caption{{RRS remanent fraction} as a function of single-trap {remanent fraction}, $\eta_{st}$, and VSA feed column remanent fraction, $\eta_{feed}$, {assuming a VSA feed cycle time of $60\,$min and a purge flow fraction of 10\%, but ignoring the intrinsic radon activity introduced by the trap. The efficacy is independent of the number of inlet radon atoms from the detector and does not have an explicit dependence on the detector parameters or the adsorptive properties of the trap. The white point indicates about 52\%  {RRS remanent fraction} as a result of a  single trap with 10\% {remanent fraction}  in the feedback loop of a feed column with remanent fraction of 90\%.}}
\label{f12a}
\end{figure}

{For illustration purposes, let us continue with the example from  Sec.~\ref{subsubsection-quasi-ideal}.} { We still neglect the intrinsic activity of the charcoal, and we still assume the VSA feed cycle time is $60\,$min, and the purge flow fraction to be 10\%. But now we integrate a single trap with a modest {remanent fraction} of 10\% in the feedback loop of the VSA with a feed column remanent fraction of 90\%. The result of such an arrangement is shown as dotted white lines in Fig.~\ref{f12a},} {which depicts a map of the RRS {remanent fraction} as a function of single-trap {remanent fraction} and feed column remanent fraction. } {A steady state RRS {remanent fraction} of 52\% is reached. This corresponds to a reduction in the steady state output radon fraction of almost a factor of two over a VSA system without a single trap of modest {remanent fraction} (see Fig.~\ref{f55} for comparison).}
Thus it appears that introducing a single trap, even with modest {remanent fraction} in the feedback loop of a VSA system\footnote{{Although a 10\% single-trap appears to have low remanent fraction, it is important to realize that the majority of the radon in the feedback loop goes through this trap many times.}} seems feasible if the intrinsic radon activity of the activated charcoal can be ignored.

\begin{figure}[ht]
\includegraphics[scale=0.49]{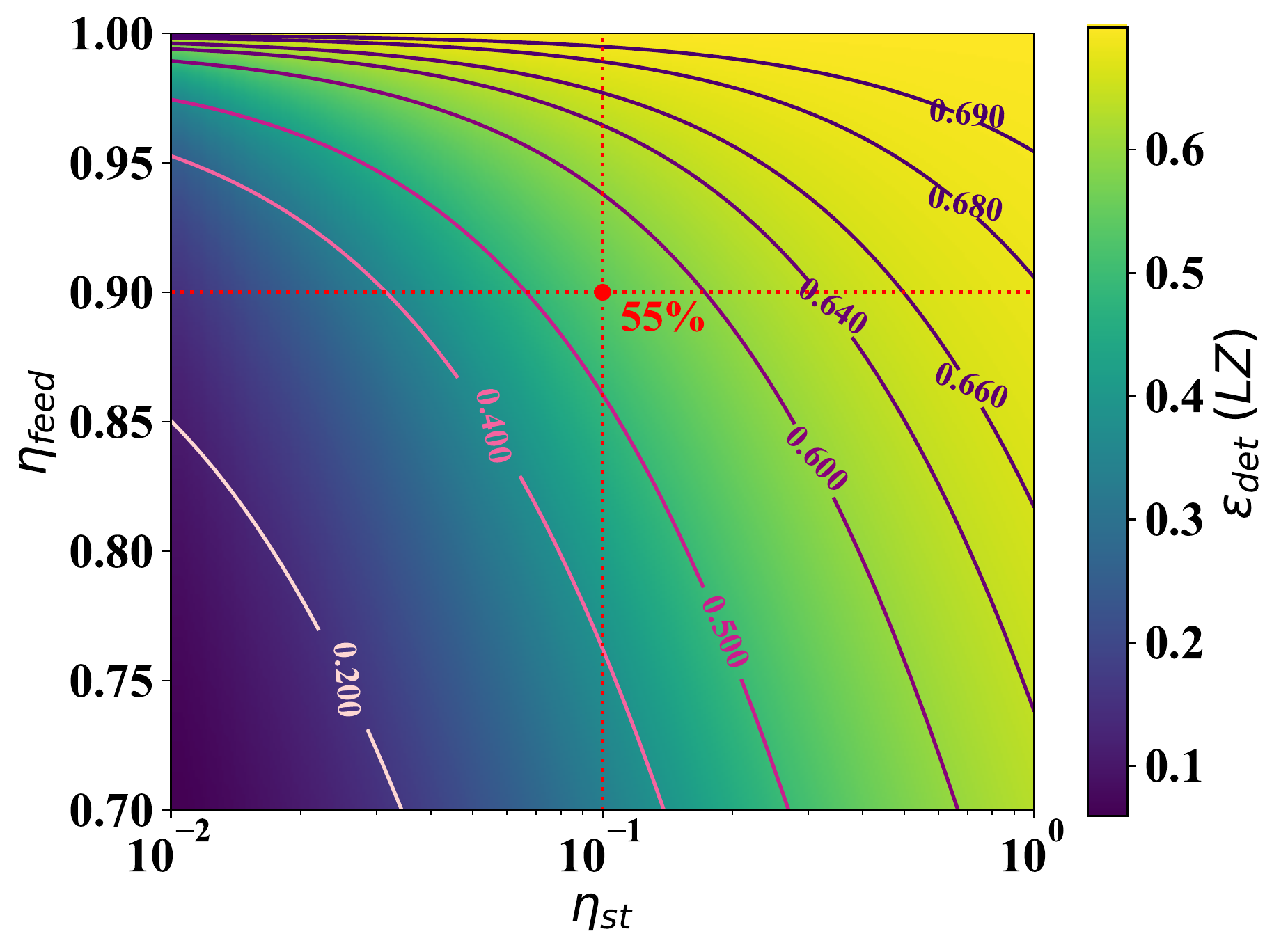}
\centering
\caption{Steady state radon reduction efficacy in a TPC detector such as LZ ($F = 500\,SLPM$ and $M = 10,000\,kg$) with a swing adsorption RRS in the main circulation path, as a function of {single-trap {remanent fraction}, $\eta_{st}$, and VSA feed column remanent fraction, $\eta_{feed}$, assuming a VSA feed cycle time of $60\,$min, a purge flow fraction of 10\%, {and a negligible RRS intrinsic activity,
$S_{RRS}$.} The efficacy is independent of the number of inlet radon atoms from the detector, and does not have explicit dependence on the detector parameters or adsorptive properties of the trap.}
The red point indicates about 55\% radon reduction efficacy in the TPC detector with a RRS {remanent fraction} of 52\%.}
\label{f12}
\end{figure}

Based on Eq.~(\ref{e7b}), radon reduction within a TPC detector, with a RRS of negligible intrinsic activity (i.e. $S_{RRS}=0$) can be computed for a given {remanent fraction} of the RRS {and detector volume-exchange time.} For the 52\% {remanent fraction} considered in the example, the radon reduction efficacy within LZ  ($F = 500\,SLPM$ and $M = 10,000\,kg$) is calculated to be 55\%, which is close to the maximal 70\% achievable with a perfect RRS system. Combining the results from Fig.~\ref{f12a} with Eq.~(\ref{e7b}), the steady state radon reduction efficacy in the LZ detector with a swing adsorption RRS in the main circulation path, as a function of {the single trap {remanent fraction} $\eta_{st}$} {and VSA feed remanent fraction $\eta_{feed}$} is shown in Fig.~\ref{f12}.

\subsubsection{Swing Adsorption RRS with non-zero Intrinsic Activity}
\label{subsubsection-real}

Radon levels desired in a dark matter detector {$\mathcal{O} (1\,\mu Bq/kg)$ } are about 5 orders of magnitude lower than those required in radon-reduced clean-rooms $\mathcal{O} (100\, mBq/kg)$. It is therefore not realistic to assume that the intrinsic activity of the charcoal can be ignored. {According to Eq.~(\ref{e7}),} {the radon reduction efficacy in a detector decreases if the $S_{RRS}/S_{det}$ ratio increases. To evaluate  $S_{RRS}$, the steady state radon contribution from the RRS, the radon contributions from the single trap, the feed column, and the purge column in a feed cycle need to be determined first.}

{The radon contribution of a single trap with a 10\% remanent fraction is computed in two steps. First, the mass of the single trap with  remanent fraction $\eta_{st}$ is determined from Eq.~(\ref{Ae2b}).} {Then, the radon contribution of the single trap of mass $m_{st}=|(\tau f/k_a)ln(1-\eta_{st})|$ in a feed cycle time, according to Eq.~(\ref{Ae6}), is}
\begin{equation}
   N_{st}=A_{st}t_{feed} = \frac{s_of\tau}{k_a}\left(1-e^{-\frac{mk_a}{f\tau }}\right)t_{feed},
\label{singleTtrap_Nperfeed}
\end{equation}
{where $A_{st}$, the total radon activity  of a single trap (or a charcoal  column), is the steady state radon contribution taking into account self adsorption of radon atoms that have been emanated deeper in the column.}

{Since the flow through a feed column is not continuous, the number of radon atoms from  a feed column is not a steady state contribution. Therefore, Eq.~(\ref{singleTtrap_Nperfeed})} {needs to be modified such that the radon contribution from a feed column in a feed cycle is expressed as}
\begin{equation}
    N_{feed}= \int_{0}^{t_{feed}} dt \int_{0}^{t_b} dt' \frac{s_om}{t_b}e^{-\frac{t'}{\tau }}H(t-t') = \int_{0}^{t_{feed}} dt \int_{0}^{t} dt' \frac{s_om}{t_b}e^{-\frac{t'}{\tau }},
\label{Be3}
\end{equation}
where $H(t-t')$ is the Heaviside step function, {$m$ is the mass of the feed column, and $s_o$ is specific the activity of the charcoal. Equation~(\ref{Be3})} {only includes the radon contribution from the part of the trap that had enough time to reach the outlet, and assumes that $t_b > t_{feed}$. {The VSA feed column} contribution in a single feed cycle is obtained from} Eq.~(\ref{Be3}) to be
\begin{equation}
    N_{feed} = \tau^{2}\frac{s_om}{t_b}\left[e^{-\frac{t_{feed}}{\tau }}-\left(1-\frac{t_{feed}}{\tau}\right)\right],
\label{Be4}
\end{equation}
which can be approximated to
\begin{equation}
   N_{feed} \approx \frac{s_om}{2t_b}t_{feed}^2.
\label{Be4_approx}
\end{equation}
This approximation is valid since the feed cycle time of the VSA is much shorter than the radon lifetime $(t_{feed} \ll \tau)$, and {because by ignoring the radon decay term} it will always over-estimate radon content.

For a purge fraction of 10\%, the radon breakthrough time in the feed column is about an order of magnitude greater than the breakthrough time in the purge column. The mass flow rate of the purge flow is $(1/10)$ of the feed flow, but the purge pressure is $(1/100)$ of the feed pressure, resulting into a 10 times larger volumetric flow rate {$(f\propto \varphi/P$}, where $\varphi$ is the mass flow rate, $f$ is the volumetric flow rate and P is the pressure$)$~\cite{c2}. Therefore it is assumed that all radon atoms are fully purged out in the purging stage. Consequently, the VSA purge column contribution to the feedback loop in a feed cycle time is

\begin{equation}
   {N_{purge} \approx s_omt_{feed} + \left[s_omt_{feed}- \frac{s_om}{2t_b}t_{feed}^2\right]e^{-t_{feed}/\tau}.}
\label{Be4_purge_approx}
\end{equation}
{Note that $N_{purge}$ contains both the total radon contribution in {the current} feed cycle from the {purge} column ($1^{st}$ term), and the {trapped radon atoms from the previous cycle in which that column was in the feed stage ($2^{nd}$ term).}  {This $2^{nd}$ term is given by the total radon emanation minus the fraction that escaped in the feed stage} times the exponential decay term which takes into account the decay during the feed cycle time.}

To evaluate steady state conditions for the RRS, {the dynamics of emanated radon atoms in a single cycle will be considered, analogous to the procedure in Sec.~(\ref{subsubsection-quasi-ideal}).} {The radon input from the detector, $N_{det}$, is set to zero, and only the radon contribution of the RRS itself is considered.} The evolution of emanated radon atoms after the $n^{th}$ feed of the RRS is
\begin{subequations}
\begin{equation}
    (N_{out})_{n} = (N_{in})_{n}(1-\eta_{feed}){e^{-t_{feed}/\tau}}r_{out} + N_{feed}r_{out},
\label{emanation_dynamics_out}
\end{equation}
and
\begin{equation}
\begin{split}
  (N_{in})_{n+1} = N_{st} + (1-\eta_{st})\Big[ & (N_{in})_n(1-\eta_{feed}){e^{-t_{feed}/\tau}}r_{purge} \\ & + (N_{in})_{n}\eta_{feed}e^{-t_{feed}/\tau }  \\ & +N_{feed}r_{purge} + N_{purge}\Big],
\end{split}
\label{emanation_dynamics_in}
\end{equation}
\end{subequations}
{where $N_{feed}$, $N_{purge}$, and $N_{st}$ are the radon contributions in a feed cycle time of the feed column, the purge column, and the single trap, respectively. $N_{out}$ is the radon contribution from the RRS in a feed cycle time, and $N_{in}$ is the number of radon atoms supplied to the inlet of the feed column in a feed cycle time from the RRS feedback loop. Note that in the very first cycle $(N_{in})_0$ is zero.} The steady state solution of Eqs.~(\ref{emanation_dynamics_out}) and (\ref{emanation_dynamics_in}) gives $S_{RRS}=(N_{out})_{ss}/t_{feed}$, the total radon contribution from the RRS.

Let us continue with the example illustrated in Sec.~\ref{subsubsection-quasi-ideal}, where a single trap with 10\% remanent fraction is integrated in the feedback loop of a VSA with a feed column that has a 90\% remanent fraction, assuming a VSA feed cycle time of $60\,$min and a purge flow fraction of 10\%. {We can now estimate the steady state radon contribution from the RRS for a system with a circulation flow rate of $500\;SLPM$ (LZ) and charcoal with adsorption coefficients of $500\,l/kg$ at $295\,K$ or $3,000\,l/kg$ at $190\,K$ (Saratech charcoal). Combining the RRS radon contribution with the radon emanation rate of the LZ detector ($20\,mBq$) and the LZ volume exchange time ($2.4\,days$) will yield the radon reduction efficacy in the LZ detector according to Eq.~(\ref{e7}).}

\begin{figure}[ht]
  \includegraphics[width=0.5\textwidth]{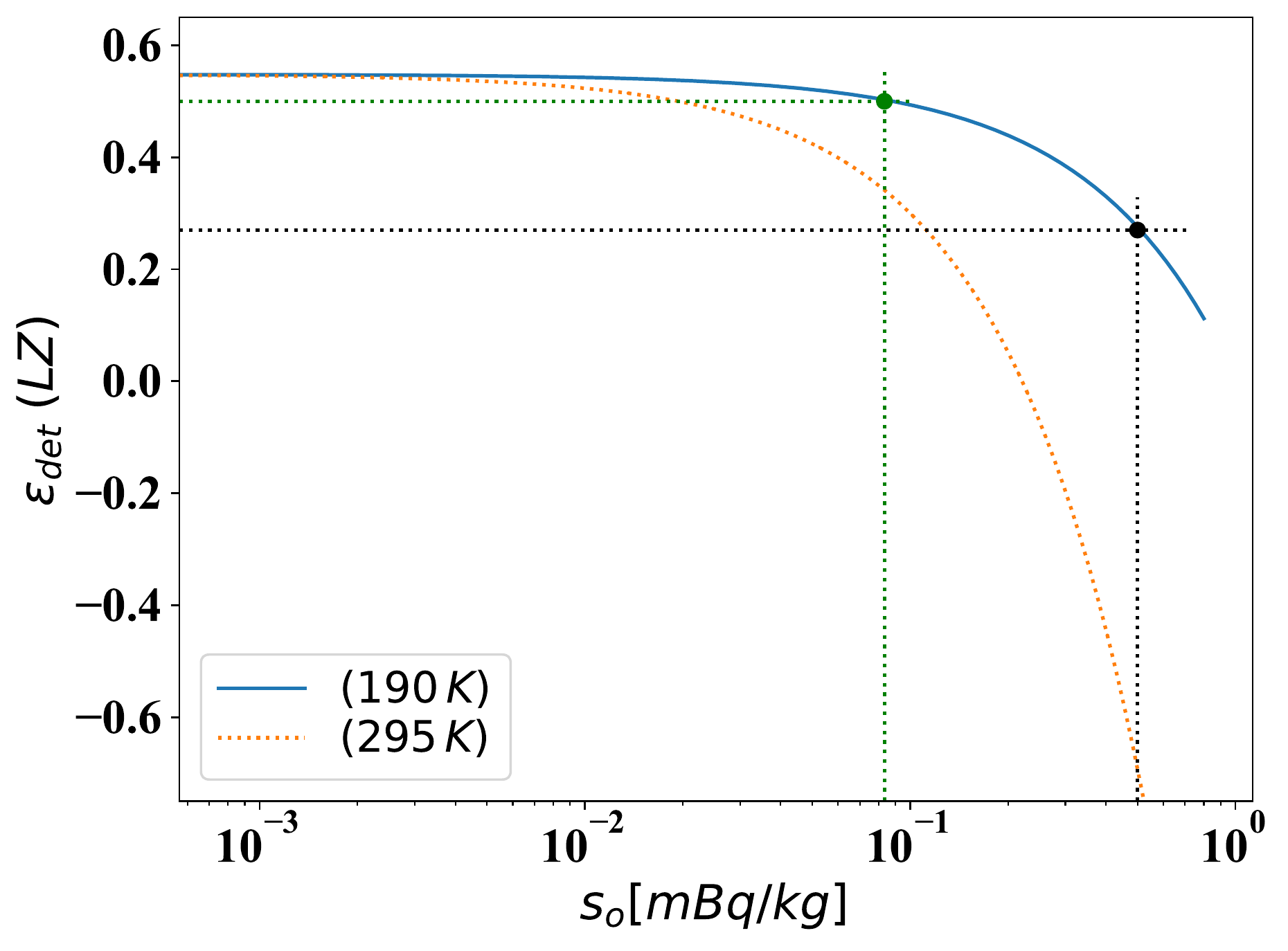}
  \centering
  \caption{Steady state radon reduction efficacy in the LZ detector with a RRS  with 52\% remnant fraction as a function of intrinsic charcoal activity, for $F = 500\,SLPM$, $M = 10,000\,kg$, $S_{det}(LZ)=20\,mBq$, a single trap in the feedback loop of a VSA with $\eta_{st}=0,1$, $\eta_{feed}=0.9$, $t_{feed}=60\,$min, and $r_{purge}=0.1$, and $k_{a}=500\,l/kg$ at $295\,K$ and  $k_{a}=3,000\,l/kg$ at $190\,K$. The black dotted lines indicate that for Saratech charcoal the radon reduction efficacy is negative (i.e. harmful) for a VSA RRS operated at room temperature, and about 27\% for a cold VSA RRS. {The green dotted lines indicate that} a specific activity  of about $0.08\,mBq/kg$ (i.e. a factor of 6 lower) is needed to achieve a factor of two radon reduction in the LZ detector, even for a cold RRS.}
  \label{realisticRRS_EFF_cold}
\end{figure}

{Figure~\ref{realisticRRS_EFF_cold} shows the radon reduction efficacy in the LZ detector as a function of the intrinsic charcoal activity. The black dotted lines {indicate} that a RRS with a charcoal activity of $0.5\,mBq/kg$ (i.e. the lowest currently available activity of Saratech charcoal)  {operated at room temperature} would be quite harmful, since it introduces more radon than it removes, making the radon reduction efficacy of the detector negative. It would require a charcoal with the same adsorption properties as Saratech, but with over an order of magnitude lower activity, to be effective at room temperature.}
The situation changes drastically, when the RRS is cooled down to $190\,K$. The adsorption coefficient of the charcoal increases by a factor of six, which reduces the mass required to maintain the same breakthrough time by a factor of six; this lowers the radon contribution from the VSA columns considerably. The {black} dotted {horizontal} line  shows that the radon reduction efficacy in the LZ detector becomes about 27\%. However, to achieve a factor of two radon reduction in the LZ detector,  charcoals with about six times lower\footnote{Note, that this factor of six is unrelated to the factor of six increase in the adsorption coefficient at $190\,K$.} intrinsic activity (i.e. $0.08\,mBq/kg$) would be necessary, {as indicated by the green dotted lines}.

\begin{figure}[ht]
\includegraphics[scale=0.5]{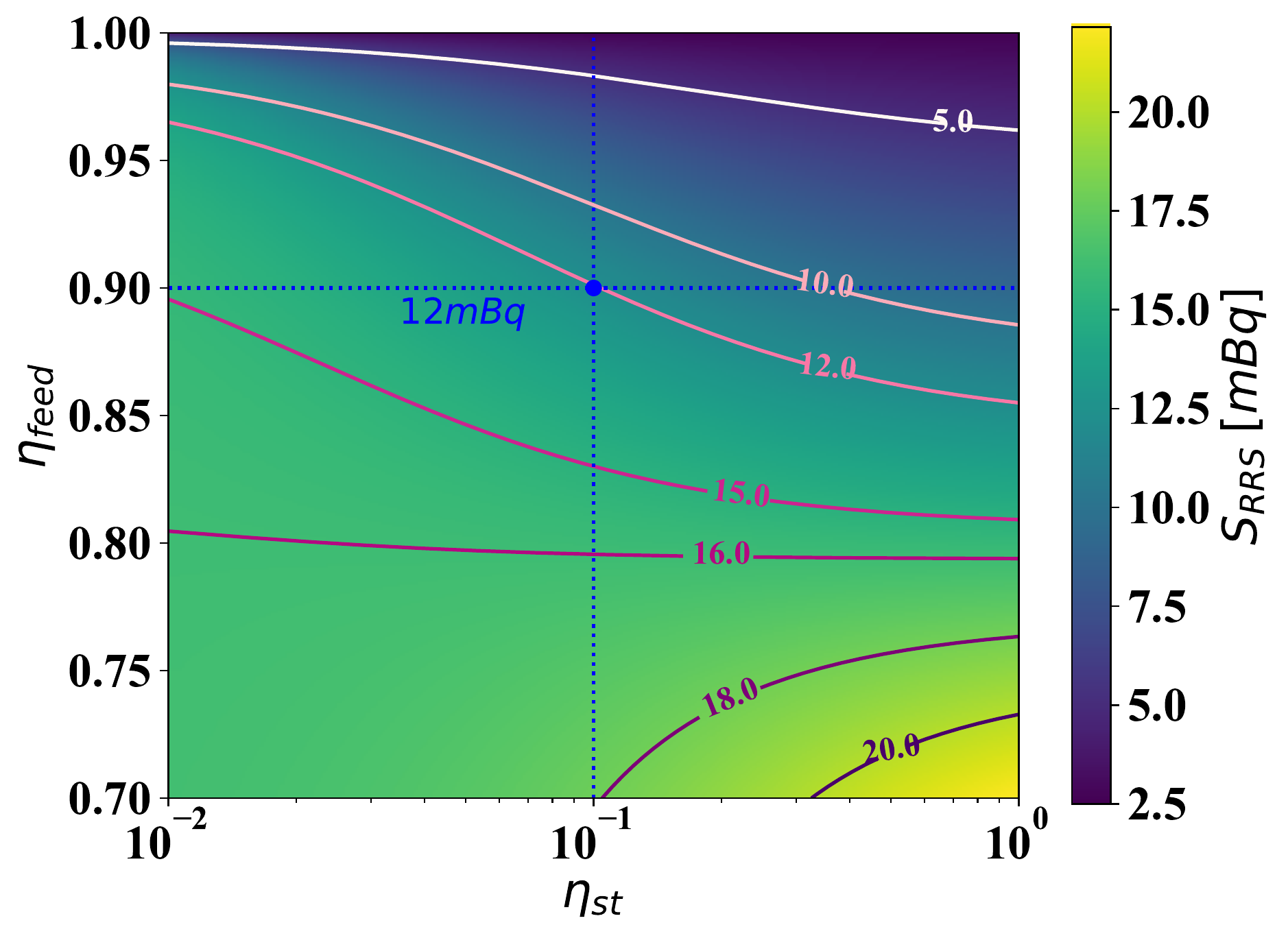}
\centering
\caption{Total radon contribution from a cold RRS as a function of single-trap remanent fraction, $\eta_{st}$, and VSA feed column remanent fraction, $\eta_{feed}$, using a charcoal with $0.5\,mBq/kg$  intrinsic activity, and assuming $t_{feed}=60\,$min,  $r_{purge}=0.1$, $F = 500\,SLPM$ and $k_{a}=3,000\,l/kg$. The blue point indicates that the radon contribution from a  single trap with 10\% remanent fraction in the feedback loop of a VSA feed column with remanent fraction of 90\% is about $12\,mBq$.}
\label{SRRS-realistic}
\end{figure}

Figure~(\ref{SRRS-realistic}) explores the radon contribution of the RRS for a broader range of values of remanent fractions in a single trap and a VSA feed column. For the example illustrated in Sec.~\ref{trap-in-loopl}, an RRS with 52\% remanent fraction, as shown in Fig.~(\ref{f12a}) and operated at $190\,K$,  results in a $12\,mBq$ steady state total radon contribution for a charcoal with $0.5\,mBq/kg$ intrinsic activity. {It further indicates that at higher single-trap remanent fractions, the single-trap contribution dominates for relatively low feed column remanent fractions.} {This may seem surprising. However, the single-trap efficacy improves by increasing its mass, which in turn increases the radon contribution from the single-trap. For a sufficiently large single-trap mass, its radon contribution starts to dominate.}

\begin{figure}[ht]
\includegraphics[scale=0.5]{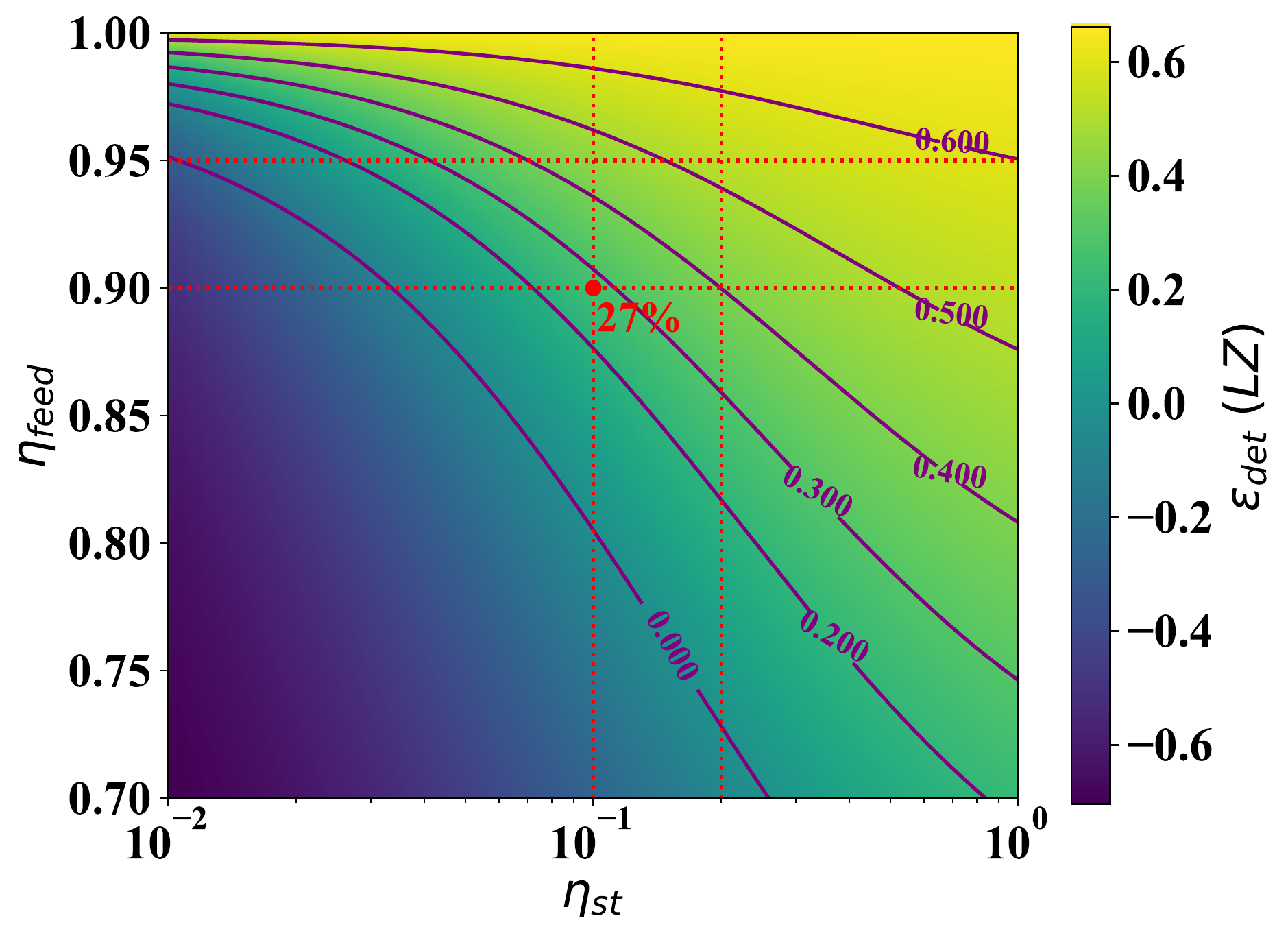}
\centering
\caption{Steady state radon reduction efficacy in the LZ detector ($F = 500\,SLPM$, $M = 10,000\,kg$, and $S_{LZ}=20\,mBq$) with a cold swing adsorption RRS in the main circulation path, as a function of single-trap {remanent fraction}, $\eta_{st}$, and VSA feed column remanent fraction, $\eta_{feed}$, assuming
$t_{feed}=60\,$min,  $r_{purge}=0.1$,  $s_{0}=0.5\,mBq/kg$, and $k_{a}=3,000\,l/kg$. The red dotted lines indicate that the radon reduction efficacy in the LZ detector is about 27\% for $\eta_{st}=0.1$, and $\eta_{feed}=0.9$, but grows to over 50\% for $\eta_{st}=0.2$, and $\eta_{feed}=0.95$.}
\label{realistic-Edet}
\end{figure}

The RRS remanent fraction map of Fig.~(\ref{f12a}) and the total radon contribution of such a cold RRS, as shown in Fig.~(\ref{SRRS-realistic}), can be combined using Eq.~(\ref{e7}) to give the steady state radon reduction efficacy. Figure~(\ref{realistic-Edet}) {shows a map of the radon reduction efficacy in the LZ detector taking into account the radon contribution from the RRS as function of single-trap remanent fraction and feed column remanent fraction. For the example of a single trap with a 10\% remanent fraction and a VSA feed column of 90\% remanent fraction, the steady state radon reduction efficacy with a cold RRS is about 27\%. However, if slightly higher remanent fractions are assumed ($\eta_{feed} = 0.95$, and $\eta_{st} = 0.2$), the resulting RRS would have a radon reduction efficacy of over 50\% in the LZ detector with existing charcoal (Saratech). As discussed in Sec.~\ref{subsubsection-quasi-ideal},} {this may not be too optimistic for a VSA system that contains Saratech charcoal. Increasing the remament fraction of the single trap from 10\% to 20\% would require to increase the trap mass from $15\,kg$ to $33\,kg$.}

\section{Conclusion}\label{section-conc}

Radon and its daughters constitute the most significant backgrounds in rare event searches since they are continuously resupplied from detector materials. Although radon screening of every single detector component is vital to reach high sensitivity for dark matter detection, it is not sufficient.
Further mitigation strategies are required that include both, in-situ hardware radon reduction and background discrimination in the analysis of the data.

The performance of charcoal-based radon reduction systems has been explored. For illustration purposes, references to the LZ detector have been made, but the general arguments and observations are not limited to one specific dark matter experiment. In-line radon reduction systems in auxiliary circulation loops, as employed in the LZ experiment, reduce radon-rich gaseous xenon from the warm components of the detector before they return radon-reduced xenon back to the main circulation loop.  This single, charcoal-based, adsorption trap approach is effective for slow circulation flow rates, but breaks down at high circulation flow rates, which are required to purify entire volumes of ton scale or larger noble-liquid detectors.
It is found that scaling up charcoal-based single-trap radon reduction systems to make them viable at the circulation flow rates of multi-ton TPC detectors is impractical even if radon emanation from charcoal is negligible.

Vacuum swing adsorption systems, which have shown great success at reducing atmospheric radon levels in clean-rooms, have clear advantages over single-trap  systems. However, they need to be modified so that they can capture and return the noble carrier gases to the purification system  through a gas feedback loop rather than releasing them into the atmosphere. The drawback of such systems is that the radon atoms become effectively trapped and can lead to many-fold higher radon concentrations in the feedback circulation loop. This can be ameliorated by introducing a modest cold single trap into the feedback loop. It allows the radon atoms to accumulate and decay in the single trap, rather than in the charcoal columns of the swing adsorption system, where some fraction can escape and be reintroduced into the TPC detector. It is found that for a VSA system with zero intrinsic activity and a feed column of 90\% remanent fraction, introduction of even a 10\% remanent fraction single trap in the feedback loop reduces the steady state output radon fraction from about 90\% to about 50\%.

While this is encouraging, it needs to be pointed out that {VSA} systems too are limited by the intrinsic radon activity of their charcoal adsorbent, particularly if they are operated at room temperature. Under these circumstances, adsorbents with more than an order of magnitude lower intrinsic radon activity than in currently available activated charcoals are required to build effective vacuum swing adsorption systems for rare event search experiments. If such VSA systems are instead cooled to about $190\,K$, this requirement relaxes drastically. Alternatively, effective VSA systems can be realized if 95\% or higher remanent fractions can be achieved. If either of these requirements can be met, vacuum swing adsorption systems may be viable options for effective radon reduction systems by the time future generation experiments are realized. Other options, not pursued here, might include radon purification in the liquid phase.

\acknowledgments

We acknowledge support of the U.S. Department of Energy (DOE) Office of Science under grant numbers DE-SC0015708 and DE-SC0019193, and under contract  number DE-AC02-76SF00515, the SLAC National Accelerator Laboratory and the University of Michigan. We would like to thank Kirill Pushkin at the University of Michigan for helpful conversations. We would also like to thank the members of the LZ collaboration for many insightful discussions.

\appendix

\section{Considerations for the Feed Cycle Time in a Swing Adsorption RRS}
\label{appendix_B}

\begin{figure}[ht]
\includegraphics[width=0.5\textwidth]{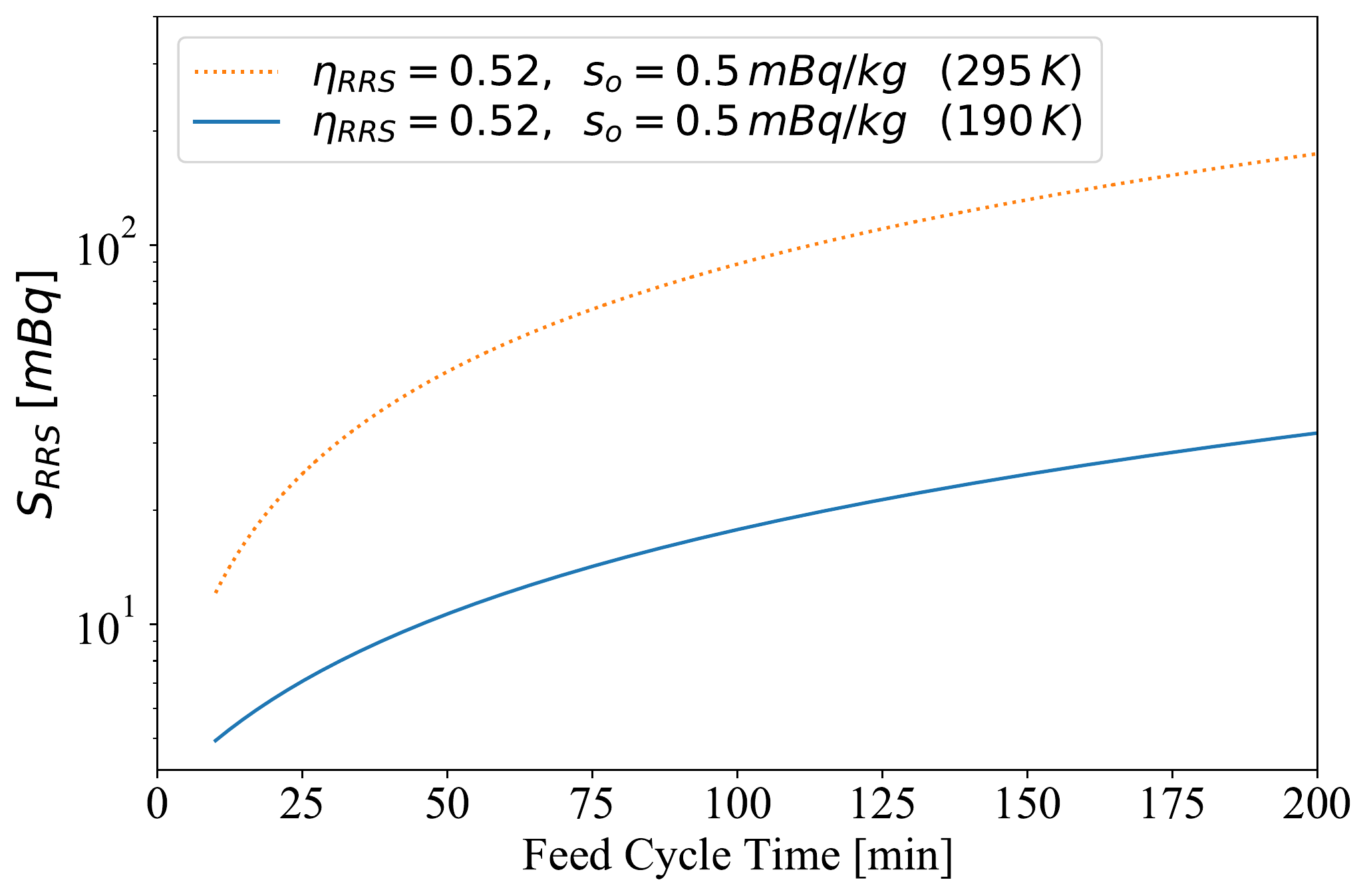}
\centering
\caption{Steady state RRS radon contribution versus feed cycle times  in $10-200\,$min range.  The parameters used are $\eta_{st} = 0.1$, $\eta_{feed} = 0.9$, and  $r_{purge}=0.1$. The dotted orange and the solid blue curves assume a RRS remanent fraction of 52\% and $s_o=0.5\,mBq/kg$ intrinsic activity of charcoal. The blue curve represents the remanent fraction of the RRS when the VSA is cooled down to $190\,K$ $(k_a=3,000\,l/kg)$, and the orange dotted curve is at a VSA operational temperature of $295\,K$ $(k_a=500\,l/kg)$.}
\label{feedCycleTimes}
\end{figure}

Throughout this document we have used a feed cycle time of $60\,$min as an example; however, this choice is not without consequence. As shown in Fig.~\ref{feedCycleTimes}, the total radon contribution from a RRS depends on the feed cycle time. When there is no radon emanation from the RRS (i.e. $S_{RRS}=0$), then a longer feed cycle time is advantageous -- radon atoms captured in the VSA's feedback loop have a 10\% chance to escape each cycle, so a longer feed cycle time provides them fewer opportunities before their eventual decay. Conversely, in the realistic case where there \itshape is \upshape emanation from the RRS, then a shorter feed cycle time is preferable.  This is a consequence of the highly-effective purge cycle leaving the column free of radon. Atoms emanated must transit to the trap's exit within the feed cycle time in order to contribute to the concentration at the system output, so the emanation contribution is proportional to $t_{feed}^2$, as shown in Eq.~(\ref{Be4_approx}).

While a shorter feed cycle time is advantageous in a system dominated by charcoal emanation, it is technically challenging to implement a system with an arbitrarily small feed cycle time. Any real system will require some finite time to evacuate the column, bringing it from forward flow pressure to purge cycle pressure, during which time the purge cycle is ineffective.
In practice this means $t_{purge} = t_{feed} - t_{pump}$, so the statement that $t_{purge}=t_{feed}$ is only true when $t_{feed} \gg t_{pump}$.   Typical VSA systems in air require tens of seconds to pump out, so this condition is achieved with feed cycle times of approximately one hour or longer.   Thus, a feed cycle time of one hour was used for the calculations in this paper.

{As discussed in Sec.~\ref{subsubsection-quasi-ideal}, } { the tight elution curves of Saratech charcoal allow  traps to have breakthrough times that are only about 50\% longer than the feed cycle time, but with negligible radon breakthrough. This leads to breakthrough times of  $t_b =  1.5\; t_{feed} =  90\,$min, and in turn, since $t_b = k_am/f$, to VSA columns that are of $\mathcal{O}({100\;kg})$  if operated at room temperature ($k_a = 500\,l/kg$), and $\mathcal{O}({20\;kg})$  if operated at $190\,K$  ($k_a = 3,000\,l/kg$), at flow rates near $f=500\;SLPM$.}

There is one additional consideration, especially for the cold VSA discussion.
In order for the purge cycle to be effective, the characteristic time that a radon atom is stuck to charcoal must be much shorter than the duration of the purge cycle. It is possible that the low temperatures discussed here will require longer swing-cycle periods as this characteristic time increases with reduced temperature.


\begin{thebibliography}{99}
\bibitem{c9} Private communications with Bl\"ucher GMBH, the German producer of the Saratech brand charcoal, revealed that the cleanliness of their charcoal could be significantly improved if they used their quartz glass reactor rather than their metal reactor for production.
\end{thebibliography}
\end{document}